\documentclass[showkeys,pdftex,nofootinbib,superscriptaddress,aps,prfluids,notitlepage]{revtex4}
\usepackage{multirow,enumerate,epsfig,graphics,amssymb,amsmath,subeqnarray,mathrsfs,color,xcolor,epstopdf}
\usepackage{natbib}
\usepackage{color,epsfig,graphics,amssymb,amsmath,subeqnarray,graphicx,amsthm,subfigure,mathrsfs}
\usepackage{mathtools,bm} 
\usepackage{rotating,mathrsfs}

\makeatletter
\def\sgn{\mathop{\operator@font sgn}}
\def\threevdots{\vbox{\baselineskip1\p@ \lineskiplimit\z@
  \kern6\p@\hbox{.}\hbox{.}\hbox{.}}}
\makeatother

\begin{document}
\title{Flutter and resonances of a flag near a free surface}
\author{J\'er\^ome Mougel}
\email{jerome.mougel@imft.fr}
\affiliation{Institut de M\'ecanique des Fluides de Toulouse (IMFT),Universit\'e de Toulouse, CNRS, Toulouse, France}
\author{S\'ebastien Michelin}
\email{sebastien.michelin@ladhyx.polytechnique.fr}
\affiliation{LadHyX -- D\'epartement de M\'ecanique, CNRS -- Ecole Polytechnique, Institut Polytechnique de Paris, 91128 Palaiseau Cedex, France}

\begin{abstract}
We investigate the effects of a nearby free surface on the stability of a flexible plate in axial flow. 
Confinement by rigid boundaries is known to affect flag flutter thresholds and fluttering dynamics significantly, and this work considers the effects of a more general confinement involving a deformable free surface. 
To this end, a local linear stability is proposed for a flag in axial uniform flow and parallel to a free surface, using one-dimensional beam and potential flow models to revisit this classical fluid-structure interaction problem.
The physical behaviour of the confining free surface is characterized by the Froude number, corresponding to the ratio of the incoming flow velocity to that of the gravity waves. After presenting the simplified limit of infinite span (i.e. two-dimensional problem), the results are generalized to include finite-span and lateral confinement effects. In both cases, three unstable regimes are identified for varying Froude number.  
Rigidly-confined flutter is observed for low Froude number, i.e. when the free surface behaves as a rigid wall, and is equivalent to the classical problem of the confined flag. When the flow and wave velocities are comparable, a new instability is observed before the onset of flutter (i.e. at lower reduced flow speed) and results from the resonance of a structural bending wave and one of the fundamental modes of surface gravity waves. Finally, for large Froude number (low effect of gravity), flutter is observed with significant but passive deformation of the free surface in response of the flag's displacement. 

\end{abstract}
\maketitle

\section{Introduction}
The flow-induced flapping of a flexible plate in axial flows, also known as ``flag flutter", is a canonical and fundamental example of fluid-structure interaction, where the destabilizing effect of the fluid forces can overcome the plate's resistance to deformation and lead to spontaneous and self-sustained flapping above a critical flow velocity. 
 Motivated by applications ranging from fundamental aerodynamics to engineering problems, biomedical applications, acoustics or flow control, the detailed analysis of flag flutter has been central to many different studies, either theoretical, numerical or experimental over the last two decades~\cite{paidoussis2004, shelley2011}. 
Determining velocity thresholds (i.e. velocities above which the flat equilibrium becomes unstable and the plate starts flapping spontaneously) and the influence of the flag's inertia and aspect ratios is key to many such studies~\cite{eloy2007}. Beyond their fundamental interest for understanding the transition to self-sustained flapping, flutter thresholds are also crucial in most applications, whether to estimate proper operating range for flow-energy harvesters~\cite{giacomello2011, doare2011, michelin2013}, to avoid jamming in paper industry~\cite{watanabe2002exp, watanabe2002theo}, to dimension cooling systems appropriately~\cite{hidalgo2010, shoele2014} or to prevent and treat snoring in human patients~\cite{huang1995}. 
 
 In most of these applications, confinement of the fluid and flag displacements by rigid boundaries is significant, and the fundamental role of such confinement has recently been the focus of specific studies on snoring~\cite{auregan1995, balint2005, howell2009}, heat transfert devices~\cite{shoele2014, shoele2016b}, or paper manufacturing~\cite{wu2005}. 
Lateral confinement (i.e. in the direction of the flapping motion) was thus found to modify flutter thresholds as well as fluttering frequencies and mode shapes. Specifically, lateral confinement was commonly observed to destabilize the flag~\cite{dessi2015,alben2015,doare2012,guo2000}, an effect associated with an increased added inertia of the flag~\cite{jaiman2014}, as the result of the proximity of a rigid boundary~\cite{brennen1982}. 
This effect was most pronounced for heavier flags, while lateral confinement was found to have little influence on stability of lighter structures~\cite{alben2015}. The same trends also hold for span-wise confinement~\cite{doare2011num}.  
Most of the analytical and numerical studies were performed in the inviscid limit, yet confinement effects in viscous cases essentially lead to similar trends~\cite{shoele2016b,cisonni2017}. Additionally, experimental and numerical studies of span-wise confinement demonstrated the difficulty of exploiting such confinement to analyse two-dimensional flag flapping in experiments~\cite{doare2011num,doare2011exp}. 

The previous considerations apply only to non-deforming boundaries, i.e. rigid walls. In contrast, the present study considers the effect of confinement by a free surface which may deform under the effect of the fluctuating pressure generated by the flapping structure. Such analysis is necessary to understand flag-current-free surface coupling and its relevance for the design of flexible energy harvesters intended for ocean or river purposes~\cite{giacomello2011, trasch2018, muriel2018}, the performance prediction of recent self-powered water-pump devices based on flexible structures~\cite{muriel2016}, or the analysis of the effect of a current on wave energy harvesters~\cite{alam2012} and wave breakers~\cite{cho2000}.

This configuration also presents much fundamental interest to explore the potential coupling of the flag dynamics to the flow-induced deformation of the free surface. This is expected to lead to new physical phenomena, in comparison with purely rigid confinement, since a free surface is also susceptible to develop and carry neutrally-stable waves, therefore possessing its own dynamics. Such coupling between a flag and other deforming boundaries represents a wider class of problems which includes the interaction of multiple flags: parallel arrangement of the flags is generally found to reduce flutter thresholds and significantly complexify fluttering dynamics \cite{schouveiler2009, mougel2016}. 
In that case, however, the physical mechanism at the root of the instability essentially remains the same as for the single-flag configuration owing to the specificity and symmetry of the coupling considered, namely flag-fluid-flag coupling. 

Recently, several asymmetric couplings of a flapping flag with another dynamical system have been considered and may offer some insight on the accessible dynamics. Fluid-solid-electric coupling in a piezoelectric flag have thus been considered and result in the lock-in of the flapping motion onto the natural frequency of the output electric circuit, which significantly affects their energy harvesting efficiency~\cite{xia2015}. Furthermore, continuous circuit models of piezoelectric flags showed some further coupling and organization of the energy transport along the solid and electric systems~\cite{xia2017}. The coupling of a flag to a mechanical resonant system through the rotation of its mast also displayed the possibility of lock-in and increased energy transfer~\cite{virot2016}. It is therefore expected that flag-current-free surface coupling may lead to interesting dynamics which are the focus of the present study. 

To analyse the coupling of the flag dynamics with free surface wave dynamics, we propose here a local stability analysis of a flag immersed below a free surface and exposed to a steady horizontal current. By focusing on the propagation of waves along a flag of infinite length, this idealized configuration offers a simplified framework to characterize flag-current-free surface coupling, to quantify free surface effects on flag flutter thresholds and to investigate potential free surface-induced instabilities. {Note however that the present model is not able to capture finite-length effects of the flag (including the effect of the wake shedded at its trailing edge), are beyond the scope of the present paper. } 
 
The paper is organized as follows. Section~\ref{sec:pb_setting} first introduces the model problem considered and obtains the dimensional and non-dimensional forms of the governing equations for the flow, flag and free surface in the general case. The large-span limit, {which results in a two-dimensional problem,} is then investigated in details in Section~\ref{sec:largespan} and allows to characterize the different regimes of coupling. Section~\ref{sec:finite} then generalizes these results to a flag of finite span {(but rigid in the span-wise direction) immersed in a free surface channel of finite width}, thereby demonstrating the generality of the results and analyzing the specificity introduced by the span-wise extent and confinement of the problem. Finally, the main findings of the paper are summarized in Section~\ref{sec:conclusions} which also presents some perspectives for future work.

\section{Theoretical modelling}
\label{sec:pb_setting}
\subsection{Flag model}
\begin{figure}
\centering
\begin{tabular}{ccc}
\includegraphics[width=6.5cm, trim = 0cm 0cm 0cm 0cm, clip]{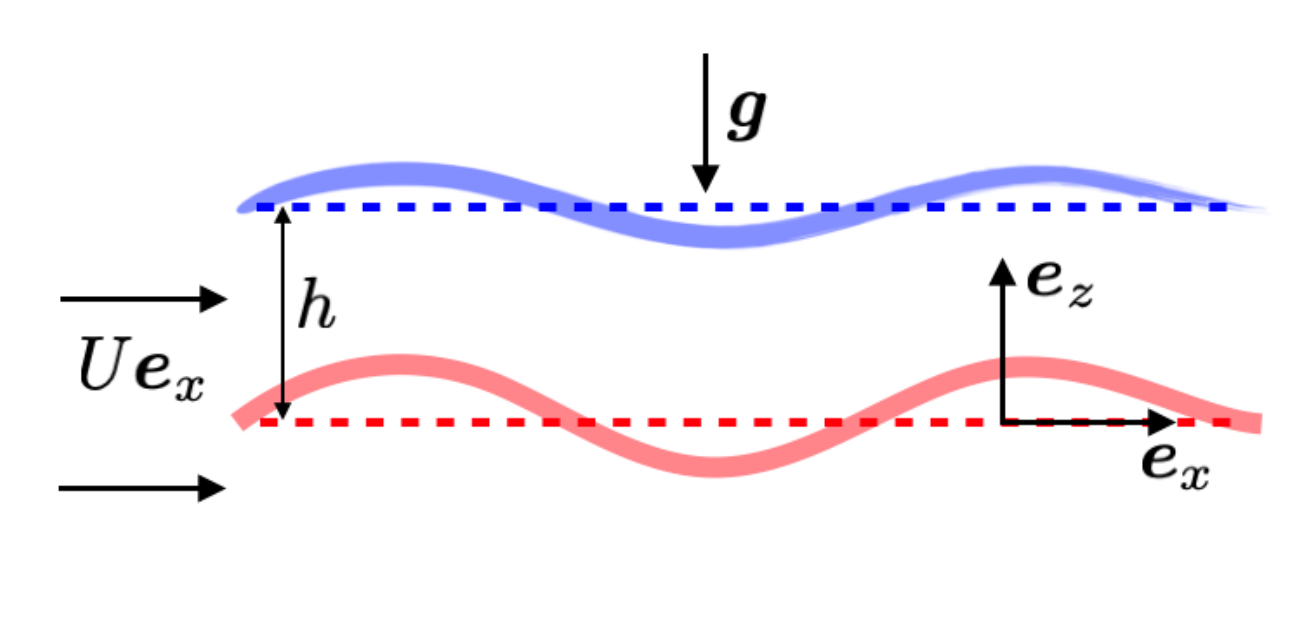} & &
\includegraphics[width=3cm, trim = 0cm 0cm 0cm 0cm, clip]{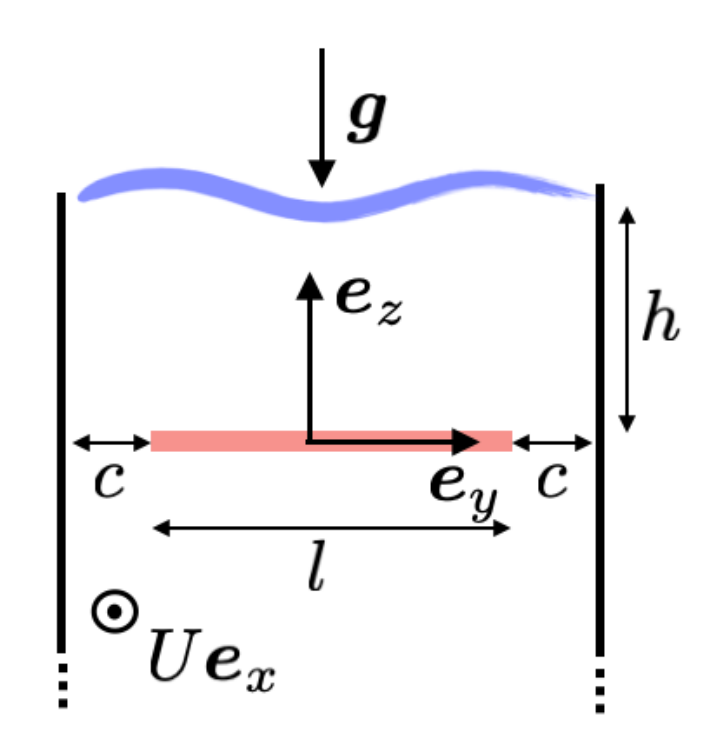}
\end{tabular}
\caption{\label{fig:sketch} Geometric description of the problem.}
\end{figure}

To analyse the local stability problem for a coupled flag and free surface, an infinitely long flexible plate with span $l$ is considered, which is {immersed horizontally} in a uniform axial flow of velocity $U$ and fluid density $\rho$, and at a distance $h$ below a free surface. This flexible plate, referred to as "flag" in the following, is confined by rigid walls in the span-wise direction located at a distance $c$ from its side edges (Figure~\ref{fig:sketch}). In the following, viscous effects are neglected in front of fluid and solid inertia (i.e. the Reynolds number is large), so that the fluid forces are dominated by pressure contributions. Physically, viscous effects are concentrated within thin boundary layers on the flag and marginally affect the flow field, except at the downstream edge where they separate into free shear layers. In the present local stability framework, modeling of such separation is abscent (the flag is considered effectively of infinite streamwise extent).

For simplicity, we assume that the flag is significantly more rigid in the span-wise direction than along the direction of the flow, an assumption commonly performed in the literature that is based upon experimental observations, and that greatly simplifies the analysis by focusing on unidimensional deformation of the structure~\cite{eloy2007}. Additionally, this assumption is relevant for energy harvesting devices in water \cite{giacomello2011, trasch2018}, and indeed adding heavy rigid stripes along the flag's span, heavy, but highly flexible, plates can be built, with significantly lower instability thresholds, which extends the range of operability of such devices~\cite{shelley2005}. 

Assuming that the thickness of the flag is much smaller than any other dimension in the problem, the displacement of the flag from its straight horizontal position is entirely determined by the vertical position of its centerline $\eta_f(x,t)$. In the following, we focus on the linear dynamics (small $\eta_f$), and consider   a one-dimensional Euler-Bernoulli beam model, so that the linearized beam dynamics is described by
\begin{equation}
\label{eq:beam}
\mu \frac{\partial^2 \eta_f}{\partial t^2} + B  \frac{\partial^4 \eta_f}{\partial x^4}  =  -\frac{1}{l} \int_{-l/2}^{l/2}{[p] dy},
\end{equation}
with {$\mu$ the mass per unit surface of the flag,} $B$ its flexural rigidity and  $[p](x,y,t)$ the pressure jump across the flag obtained at leading order as $[p] = p(x,y,z=0^+,t)-p(x,y,z=0^-,t)$.

In a \emph{global} stability analysis of the classical cantilevered flag problem,  Eq.~\eqref{eq:beam} would have to be complemented by clamped-free boundary conditions at the flag leading and trailing edges. This study focuses instead exclusively on the \emph{local} stability problem considering the linearized time-dependent evolution of wave-like perturbations to the steady and straight equilibrium of an infinitely long flag. 
In the unbounded two-dimensional case (i.e. infinite span), long waves (i.e. with small wave-number) are  always unstable for any non-zero velocity (as mentioned by~\cite{eloy2007} and established in the following): physically, the restoring effect of the flag's bending rigidity is small for such long waves, and it is thus unable to counter-act the destabilizing effect of the fluid pressure forcing. 
A rough, but simple model to estimate flutter thresholds of a finite-length flag traditionally consists in restricting  the present local stability analysis to perturbation wavelengths of the order of, or smaller than, the flag's length (thus considering that this is the longest wave length that can be supported by the flag). Despite its simplicity, {such local model (which therefore ignores the effect of the wake)} can predict experimental flutter thresholds in a satisfactory way (e.g.~\cite{shelley2005, jia2007, schouveiler2009}). In this study, we adopt this approach to provide insight on the different physical couplings involved in the fluid-flag-free surface interaction. {Note however that the flag's wake may significantly influence its  dynamics and flutter thresholds \cite{kornecki1976, tang2008}.}

\subsection{Inviscid fluid model}
We briefly outline now the three-dimensional incompressible potential flow model adopted here for the fluid flow around the flag. Decomposing the flow field as $\mathbf{u}=U\mathbf{e}_x+\nabla\phi$, the perturbation potential $\phi$ must satisfy Laplace's equation both above ($0<z<h$) and below the flag ($z<0$):
\begin{equation}
\label{eq:laplace}
\Delta \phi =0.
\end{equation}
Once $\phi$ is known, the pressure field is obtained from the linearized unsteady Bernoulli equation,
\begin{align}
\label{eq:bernoulli}
p = -\rho \left( \frac{\partial }{\partial t}  + U  \frac{\partial }{\partial x} \right) \phi. 
\end{align}

\subsection{Boundary conditions}
The free surface elevation relative to its equilibrium position at $z=h$ is noted $\eta_g(x,y,t)$. The linearized dynamic (pressure continuity) and kinematic boundary conditions (impermeability) at the free surface respectively read
\begin{align}
\label{eq:dyn_fs}
p &= \rho g \eta_g      \hspace{2.7cm}  \text{for} \ z=h, \\
\label{eq:kin_fs} 
\frac{\partial \phi}{\partial z} &= \left( \frac{\partial }{\partial t}  + U  \frac{\partial }{\partial x} \right) \eta_g  \ \ \ \  \  \ \  \text{for}  \ z=h,
\end{align}
with $g$ the acceleration of gravity.
In a similar fashion, the linearized kinematic boundary condition on the flag provides 
\begin{align}
\label{eq:kin_flag}
\frac{\partial \phi}{\partial z} &= \left( \frac{\partial }{\partial t}  + U  \frac{\partial }{\partial x} \right) \eta_f  \ \ \ \    \text{for} \ z=0, \   \text{and} \ |y|<l/2, 
\end{align}
and span-wise confinement further imposes the impermeability of lateral walls
\begin{align}
\label{eq:lateral}
\frac{\partial \phi}{\partial y} &= 0   \qquad\text{for} \ |y|=l/2 + c.
\end{align}
Considering that the depth of the fluid domain is much larger than any other dimension of the problem (an assumption that will be relaxed in Section~\ref{sec:finite}), $\phi$ must finally satisfy
\begin{align}
\label{eq:infinity}
  \phi &= 0   \qquad      \text{for} \  z \rightarrow -\infty.
\end{align}
Equations~\eqref{eq:beam}-\eqref{eq:infinity} determine the linear dynamics of the fully-coupled fluid-structure interaction problem. 

\subsection{Normal mode decomposition}
Taking advantage of the problem invariance in the streamwise direction, a normal mode decomposition into wave-like solutions is considered
\begin{equation}
\left[\begin{array}{c}p(x,y,z,t)\\\phi(x,y,z,t)\\\eta_f(x,t)\\\eta_g(x,y,t)\end{array}\right]=\left[\begin{array}{c}\tilde{p}(y,z)\\\tilde\phi(y,z)\\ A_f\\ A_g(y)\end{array}\right]\times\mathrm{e}^{\mathrm{i}(kx-\omega t)}
\end{equation}
with $k$ and $\omega$ corresponding to the wave number and complex frequency of the mode, respectively. We focus on the temporal stability of spatially-periodic waves of wavelength $2 \pi/k$: for a given real $k$, we determine the corresponding frequency $\omega_r = \mathrm{Re}(\omega)$ and growth rate $\omega_i = \mathrm{Im}(\omega)$ as a function of the other physical parameters of the problem. A mode of given $k$ is thus unstable for given parameter values if the corresponding growth rate is strictly positive, $\omega_i>0$.

\subsection{Non-dimensionalisation}

In the following, we use $1/k$, $1/(kU)$ and $\rho/k^3$ as reference length, time and mass, and we rescale the span-wise coordinate $y$ in units of $l$.  {All variables are made non-dimensional using these reference scales in the following.} The dispersion relation {for the non-dimensional frequency $\omega$ } is then obtained from {the non-dimensional forms of} Eqs.~\eqref{eq:beam} and~\eqref{eq:bernoulli} as
\begin{equation}
\label{eq:D}
-\omega^2 +\frac{1}{{U^*}^2}  - 2 M^* (\omega -1)^2 f(\omega)=0,
\end{equation}
where $f$ is the normalized fluid loading function
\begin{equation}
\label{eq:f} 
f(\omega) = -\frac{1}{2}  \int_{-1/2}^{1/2}{[\phi_r(y, z=0^+) -  \phi_r(y, z=0^-)] dy},
\end{equation}
and the non-dimensionalized reduced velocity potential defined as ${\phi_r=\tilde\phi /[-i(\omega -1)A_f]}$ {satisfies the following set of non-dimensional equations obtained from Eqs.~\eqref{eq:laplace}-\eqref{eq:infinity} as}
\begin{align}
\label{eq:laplace_phir}
\frac{1}{{l^*}^2}\frac{\partial^2 \phi_r}{\partial y^2}&+ \frac{\partial^2 \phi_r}{\partial z^2} = \phi_r, \\
\label{eq:freesurface_phir}
\frac{\partial \phi_r }{\partial z} &= \mbox{Fr}^2 (\omega -1)^2 \phi_r, \ \ \ \ \ \ \ \  \text{for} \ z=h^*, \\
\label{eq:flag_phir}
\frac{\partial \phi_r}{\partial z} &=   1,  \hspace{2.9cm}    \text{for} \ z=0 \   \text{and} \ |y|<1/2, \\
\label{eq:spanwall_phir}
{\frac{1}{l^*}} \frac{\partial \phi_r}{\partial y} &= 0,  \hspace{2.9cm}     \text{for} \  |y|=1/2 + c^*, \\
\label{eq:bottomwall_phir}
\phi_ r&= 0,  \hspace{2.9cm}     \text{for} \  z\rightarrow-\infty.
\end{align}
{Note that $\phi_r$ corresponds to a velocity potential rescaled by flag velocity in the frame moving with the uniform flow. This variable is  introduced for convenience, and allows for instance to obtain a set of equations independent of $\omega$ in the small $\mbox{Fr}$ limit. }

The problem above is governed by six non-dimensional parameters, namely
\begin{align}
\label{eq:dimensionless_numbers}
M^*&=  \frac{\rho}{\mu k} ,\hspace{0.5cm} U^*=  U \sqrt{\frac{\mu}{B k^2}}, \hspace{0.5cm} \mbox{Fr} = U \sqrt{\frac{k}{g}},  \\
\label{eq:dimensionless_numbers2}
 h^* &= kh,  \hspace{0.5cm} l^* = kl, \hspace{0.5cm} c^* = \frac{c}{l}\cdot
\end{align}
 $M^*$ is the fluid-solid mass ratio and $U^*$, the reduced velocity, measures the relative magnitude of the fluid forces and of the internal elastic restoring force associated with the bending stiffness. The Froude number, $\mbox{Fr}$, is the ratio of the incoming flow velocity and of the typical free surface wave speed. Finally, $h^*$, $l^*$ and $c^*$ are three geometric parameters that characterize respectively the immersion depth of the flag, its length in the span-wise direction and the lateral confinement. 

Equations \eqref{eq:laplace_phir}-\eqref{eq:bottomwall_phir} formulates a non-linear eigenvalue problem for $\omega$, which is solved using an iterative Newton-Raphson method, thus obtaining $\omega$ as a function of the non-dimensional parameters defined in Eqs.~\eqref{eq:dimensionless_numbers}--\eqref{eq:dimensionless_numbers2}.

\begin{figure}
\centering
\begin{tabular}{cc}
\includegraphics[width=5.8cm, trim = 0cm 0cm 0cm 0cm, clip]{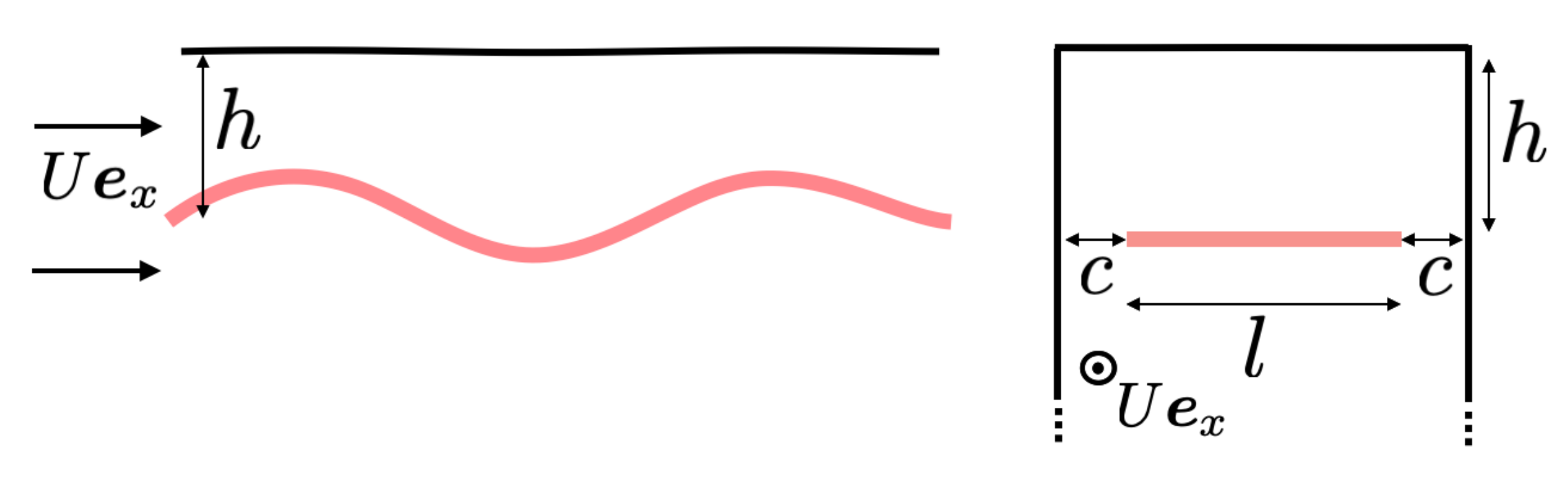} & \includegraphics[width=5.8cm, trim = 0cm 0cm 0cm 0cm, clip]{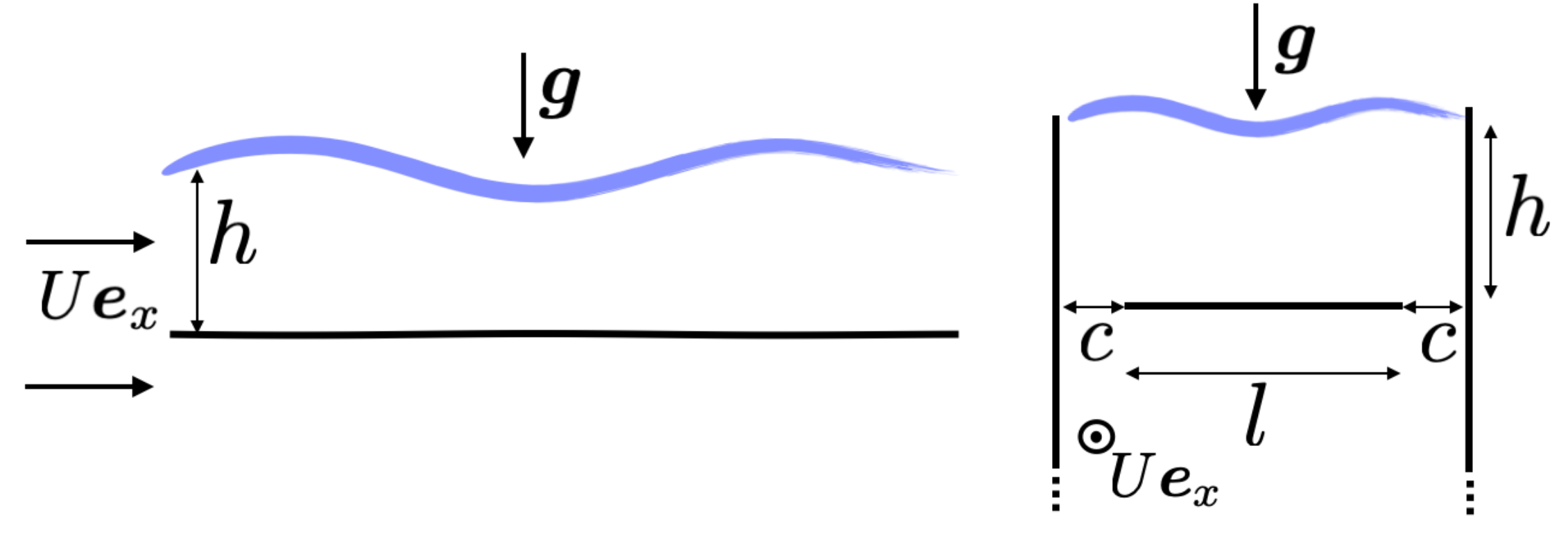}\\
$(a)$ & $(b)$
\end{tabular}
\caption{\label{fig:sketch_uncoupled} Definition of uncoupled problems. $(a)$ Uncoupled flag problem. $(b)$ Uncoupled free surface problem. }
\end{figure}

In the following, the fully coupled flag-current-free surface problem, Eqs.~\eqref{eq:D}--\eqref{eq:f}  is first analyzed by considering two simplified limits in which one of the moving boundary is fixed, namely the \emph{uncoupled flag problem} and the \emph{uncoupled free surface problem}. For the former, the top free surface is effectively replaced by a rigid wall; this limit therefore corresponds identically to a rigidly-confined flag in axial flow (Figure~\ref{fig:sketch_uncoupled}a), which has been studied by~\cite{alben2015} in the two-dimensional case. In the latter, the flag is effectively replaced by a wall, leading to a wave-current problem confined by rigid walls (Figure \ref{fig:sketch_uncoupled}b).  The motivation for introducing these uncoupled problems is two-fold, namely (i) to identify the physical nature of the coupled flag-current-free surface modes, and (ii) to provide a methodology to predict instabilities in the fully-coupled regime from inspection of the (simpler) uncoupled  problems only. 

The main objective of the present study is to quantify the effects of the free surface on the stability of the system. To that end, we mainly focus in the following on the effects of $U^*$ and $\mbox{Fr}$ on the modes' frequency, growth rate and structure. 
The Froude number $\mbox{Fr}$, Eq.~\eqref{eq:dimensionless_numbers}, compares the uniform incoming flow velocity, $U$, and the typical velocity of surface waves in the deep water limit, $\sqrt{g/k}$. 
For $\mbox{Fr}\ll 1$, surface wave propagate much faster than the flow, and this limit can thus be seen as a regime where gravity prevents any free surface deformation, which therefore behaves as a rigid boundary. {In this limit, the free surface boundary condition corresponding to equation \ref{eq:freesurface_phir} indeed degenerates to an impermeability condition.}
 Instead, when $\mbox{Fr}\gg 1$, free surface deformations are essentially advected by the flow, which behaves as a freely-deformable or \textit{soft} interface. {In this case, boundary condition \ref{eq:freesurface_phir} indeed corresponds to a zero-pressure condition at the mean free surface.} 
Varying $\mbox{Fr}$ therefore allows to span a wide variety of different physical behaviour of the free surface and resulting confinement effect on the flag.

In the remainder of the paper, we set $M^*=0.1$ and $h^*=0.5$ {and two geometric cases are explored in the following.} The large-span limit ($l^*\gg 1$) is first analysed in Section~\ref{sec:largespan}; {the two-dimensional problem obtained in this limit} is indeed amenable to analytical treatment and provides a critical insight on the coupled dynamics. A finite-span and finite-depth generalization is then investigated numerically in Section~\ref{sec:finite} {assuming no-deformations of the flag along it's span.}

\section{Large-span limit}
\label{sec:largespan}
This section focuses on the large-span limit ($l^*\gg1$), {which leads to a two-dimensional problem.}  In this limit, Eqs.~\eqref{eq:laplace_phir}-\eqref{eq:bottomwall_phir} indeed simplify into an Ordinary Differential Equation problem {($y$-derivative term is removed from Eqs.~\eqref{eq:laplace_phir} and Eqs.~\eqref{eq:spanwall_phir} can be dropped)} and can be solved explicitly for $\phi_r$ to obtain the fluid loading $f$, defined in Eq.~\eqref{eq:f}, explicitly:
\begin{equation}
\label{eq:flargespan}
f = \frac{\coth h^*}{2} \left[ \frac{1-\mbox{Fr}^2 (\omega - 1)^2 \tanh h^*}{1-\mbox{Fr}^2 (\omega - 1)^2  \coth h^*}\right] +  \frac{1}{2}\cdot
\end{equation}
After substitution into Eq.~\eqref{eq:D} the dispersion relation can be rewritten as
\begin{equation}
\label{eq:DfDg}
D(\omega) \equiv D_f(\omega)D_g(\omega) - {\cal  C}(\omega)^2=0,
\end{equation}
with
\begin{align}
\label{eq:Df}
D_f(\omega) &= \omega^2 -\frac{1}{{U^*}^{2}}  + m_a(h^*)  M^* (\omega -1)^2,\\
\label{eq:Dg}
D_g(\omega) &= -1 +  {\cal F}(h^*) \  \mbox{Fr}^2  (\omega -1)^2,
\end{align}
and
\begin{equation}
\label{eq:coupling}
{\cal{C}}(\omega)   = \frac{\sqrt{M^*}\mbox{Fr}(\omega-1)^2}{\sinh h^*},
\end{equation}
with $m_a = \coth h^* + 1$ the added mass coefficient and  ${\cal F} = \coth h^*$ a geometric factor.
Equations~\eqref{eq:Df} and \eqref{eq:Dg} can readily be identified with the large-span dispersion relations of the uncoupled flag and free surface problems, respectively. 
The dispersion relation of the fully-coupled problem, Eq.~\eqref{eq:DfDg}, can therefore be viewed as equating the product of two simplified dispersion relations corresponding to the \textit{uncoupled} cases introduced previously with a coupling term, Eq.~\eqref{eq:coupling}. 
When ${\cal{C}}(\omega)\rightarrow 0$, the eigenmodes of the fully-coupled flag/current/free surface problem identically match the four modes of the  two uncoupled problems. 
From Eq.~\eqref{eq:coupling},  this uncoupled limit arises if $h^*\gg 1$, $M^*\ll1$ or $\mbox{Fr}\ll1$, i.e. for large immersion depth, heavy flags or weak currents, respectively. 

In what follows, after providing a simple example of uncoupled solutions, we  discuss the small-coupling limit before turning to the description of unstable regimes in the fully-coupled situation.

\subsection{Uncoupled solutions}
\label{subsec:uncoupled_solutions}

\textit{Uncoupled flag problem:}
In the uncoupled flag problem, the free surface is effectively replaced by a rigid wall (Figure \ref{fig:sketch_uncoupled}a), leading to the dispersion relation ${D_f(\omega)=0}$, where the three contributions to $D_f$ in Eq.~\eqref{eq:Df} respectively correspond to flag inertia, bending rigidity, and fluid pressure. The added mass coefficient, ${m_a = \coth h^* + 1}$, includes two terms accounting for the fluid contributions on the upper part of the flag (confined fluid layer) and on the lower part of the flag (unbounded fluid layer). 
Solutions $\omega_f^{\pm}$ of the quadratic dispersion relation are structural bending waves coupled to the flow. 

\begin{figure}
\centering
\includegraphics[width=8.5cm, trim = 0cm 0cm 0cm 0cm, clip]{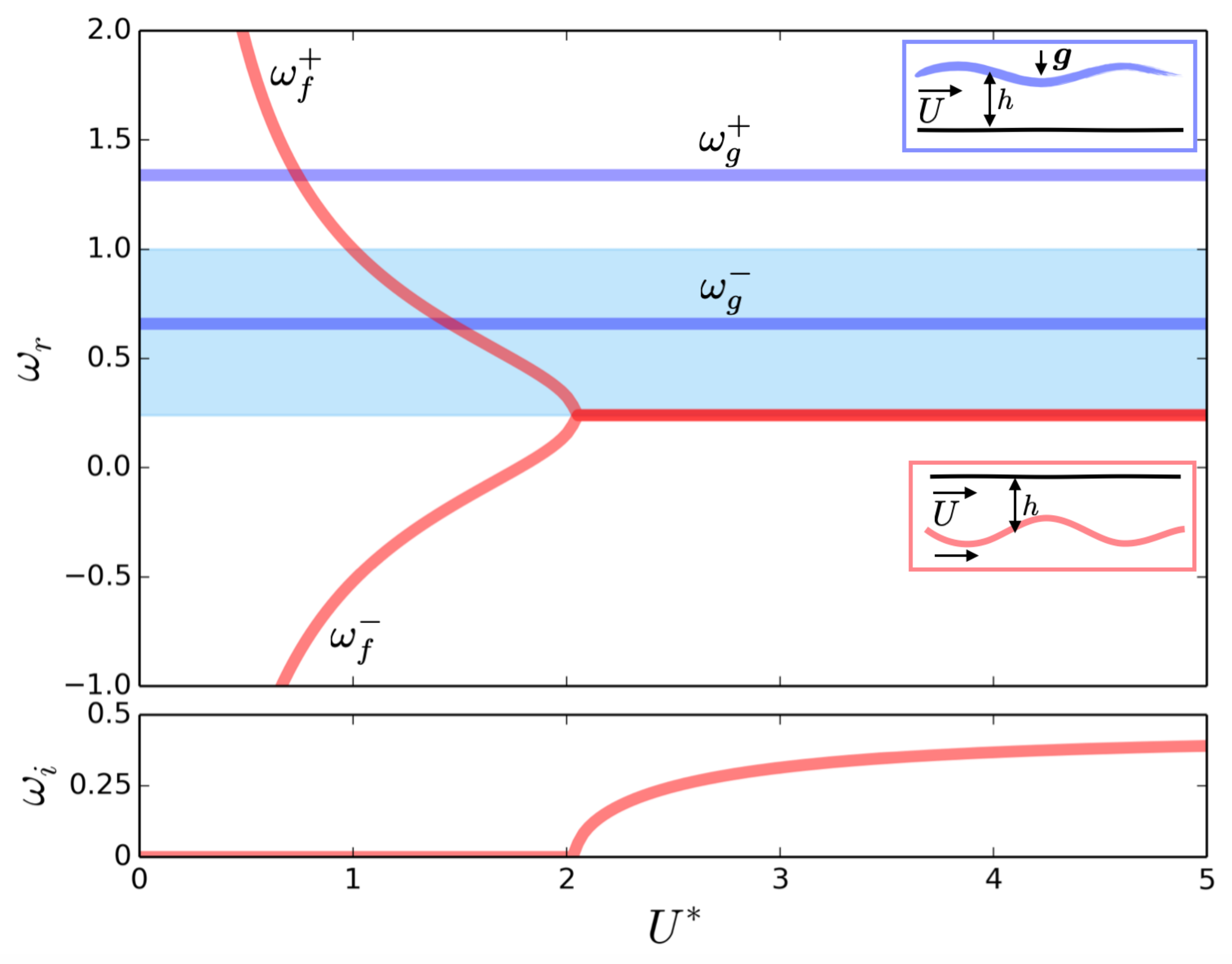} 
\caption{\label{fig:freq_uncoupled} Solutions for the uncoupled free surface problem (gravity waves, blue) and uncoupled flag waves (red) for $\mbox{Fr}=2$. The shaded blue area corresponds to the frequency range given by Eq.~\eqref{eq:instab_cond2}. Each uncoupled problem configuration is also represented schematically.}
\end{figure}

The evolution with $U^*$ of the corresponding frequencies and growth rate are shown on Figure~\ref{fig:freq_uncoupled} (red lines). For small $U^*$, the system's eigenmodes consist of two neutrally-stable waves; these merge at larger $U^*$ leading to a coupled mode flutter instability above a flutter threshold
\begin{equation}
\label{eq:Uc}
U_c^*=\sqrt{1 + \frac{1}{m_a M^*}} 
\end{equation} 
which clearly establishes the destabilizing effect of confinement (decreasing $h^*$ leads to an  increase of $m_a$)~\cite{doare2011num, alben2015, dessi2015}. Note that, recasted under dimensional form, Eq.~\eqref{eq:Uc} shows that $U_c \rightarrow 0$ in the small $k$ limit. This confirms that infinitely long flags are always unstable (as mentioned in~\cite{eloy2007}), and generalizes this result to confined flags. {Note that this is in agreement with the results of \cite{tang2008} that long flags are less stable.} \\

\noindent \textit{Uncoupled free surface problem:}
In the uncoupled free surface problem, leading to the dispersion relation  $D_g(\omega)=0$, see Eq.~\eqref{eq:Dg}, which is the two-dimensional dispersion relation of surface waves in finite-depth fluid layers in the presence of a steady current. Its two solutions, noted $\omega_g^{\pm}$, are neutrally-stable, do not depend on $U^*$ and travel respectively faster and slower than the incoming flow (Figure~\ref{fig:freq_uncoupled}). For small $\mbox{Fr}$, the latter propagates upstream.

As a result, depending on $\mbox{Fr}$, there may be particular values of $U^*$ for which a frequency of an uncoupled flag mode matches exactly that of a free surface uncoupled mode, an event referred to in the following as frequency crossing. In some circumstances, the coupling of two waves of similar frequency is known to lead to an instability of the coupled problem~\cite{cairns1979}, which motivates further analysis of the weakly-coupled limit in the vicinity of such frequency crossings.

\subsection{Weak-coupling analysis around crossings}
\label{subsec:weak_coupling}
 {We analyse the behaviour of the system for values of $U^*$ for which uncoupled solutions are close to frequency crossings, i.e. such that $\omega_g=\omega_{f}+\epsilon$, with $\epsilon \ll 1$, and focus on the weak-coupling limit assuming $\cal{C} = {\cal{O}}(\epsilon)$ (e.g. for large immersion depths $h^*$).}
We further consider a crossing occurring for values of $U^*$ smaller than the flutter threshold, such that the uncoupled solutions are neutrally-stable waves ($\omega_f$ and $\omega_g$ real). {After substitution of solutions under the form $\omega = \omega_f + \delta$, with $\delta = {\cal{O}}(\epsilon)$ and Taylor series expansion for $\epsilon\ll 1$, Eq.~\eqref{eq:DfDg} rewrites at leading order and collecting (dominant) ${\cal{O}}(\epsilon^2)$-terms}    
\begin{equation}
\delta^2 - \epsilon \delta - \gamma=0,\qquad\textrm{with    }\gamma = \frac{{{\cal{C}}(\omega_f)}^2}{\left.\frac{\partial D_g}{\partial \omega} \right|_{\omega_g} \left.\frac{\partial D_{f}}{\partial \omega} \right|_{\omega_f}}\cdot
\end{equation}
Unstable solutions for $\delta$ (i.e. complex solutions with positive imaginary part) are obtained if $|\epsilon| < 2 \sqrt{|\gamma|}$ and $\gamma<0$. The first condition indicates that frequencies of the uncoupled solutions must be close enough. The second condition equivalently writes as 
\begin{align}
\label{eq:instab_cond}
\left.\frac{\partial D_g}{\partial \omega} \right|_{\omega_g} \left.\frac{\partial D_{f}}{\partial \omega} \right|_{\omega_f} < 0.
\end{align}
Using Eqs.~\eqref{eq:Df} and \eqref{eq:Dg}, it can be shown that (at leading order)  instability is obtained if a crossing occurs within the frequency range 
\begin{equation}
\label{eq:instab_cond2}
\frac{m_a M^*}{1 + m_a M^*}< \omega < 1,
\end{equation}
with $m_a = \coth h^*+1$. 
The left and right inequalities in Eq.~\eqref{eq:instab_cond2} respectively correspond to the uncoupled flag and gravity waves. As a result, instability can only result from the coupling  with a flag wave of the surface wave slower than the flow speed (i.e. $\omega_g^-$): among the two crossings observed on Figure~\ref{fig:freq_uncoupled}, only the one involving $\omega_g^-$ (at $U^*\approx 1.5)$ can lead to an instability. Additionally, instabilities resulting from such interactions cannot appear if $\omega_g^-<0$, i.e.  $\mbox{Fr}<\sqrt{\tanh h^*}$: only supercritical cases may lead to this kind of instabilities. 

 The present analysis thus provides a predictive argument for some instabilities of the fully-coupled problem by inspection of the uncoupled problems only. 
It was adapted from the work of Cairns~\cite{cairns1979} who showed that a wave satisfying a well-defined dispersion relation $D(\omega)=0$ can be associated with energy $(\omega A^2/4) (\partial D/\partial \omega)$ with $A$ the wave amplitude, which can be interpreted as the energy to provide to the system to generate the wave from the base state. In that sense, such wave energy can be negative, meaning that exciting such waves lowers the total energy of the system. 
Building upon this work, the condition for instability in Eq.~\eqref{eq:instab_cond} can therefore be interpreted as a requirement that the two interacting waves have energies of opposite signs.  In the present case, the negative energy wave is the ``slow'' uncoupled gravity wave $\omega_g^-$. Note that such wave interactions, and hence flag/free surface instabilities, belong to class C instabilities in the three-fold classification proposed by Benjamin~\cite{benjamin1963}.
Recently, a similar analysis interpreted the formation of \textit{rotating polygons} at the interface of a swirling free surface flow in terms of wave interaction~\cite{tophoj2013}. Additionally, resonances involving such waves with energy of opposite signs have been studied in geophysics~\cite{sakai1989} and astrophysics~\cite{joarder1997}.
\begin{figure}
\centering
\begin{tabular}{cc}
\includegraphics[width=6cm, trim = 0cm 0cm 1.8cm 0cm, clip]{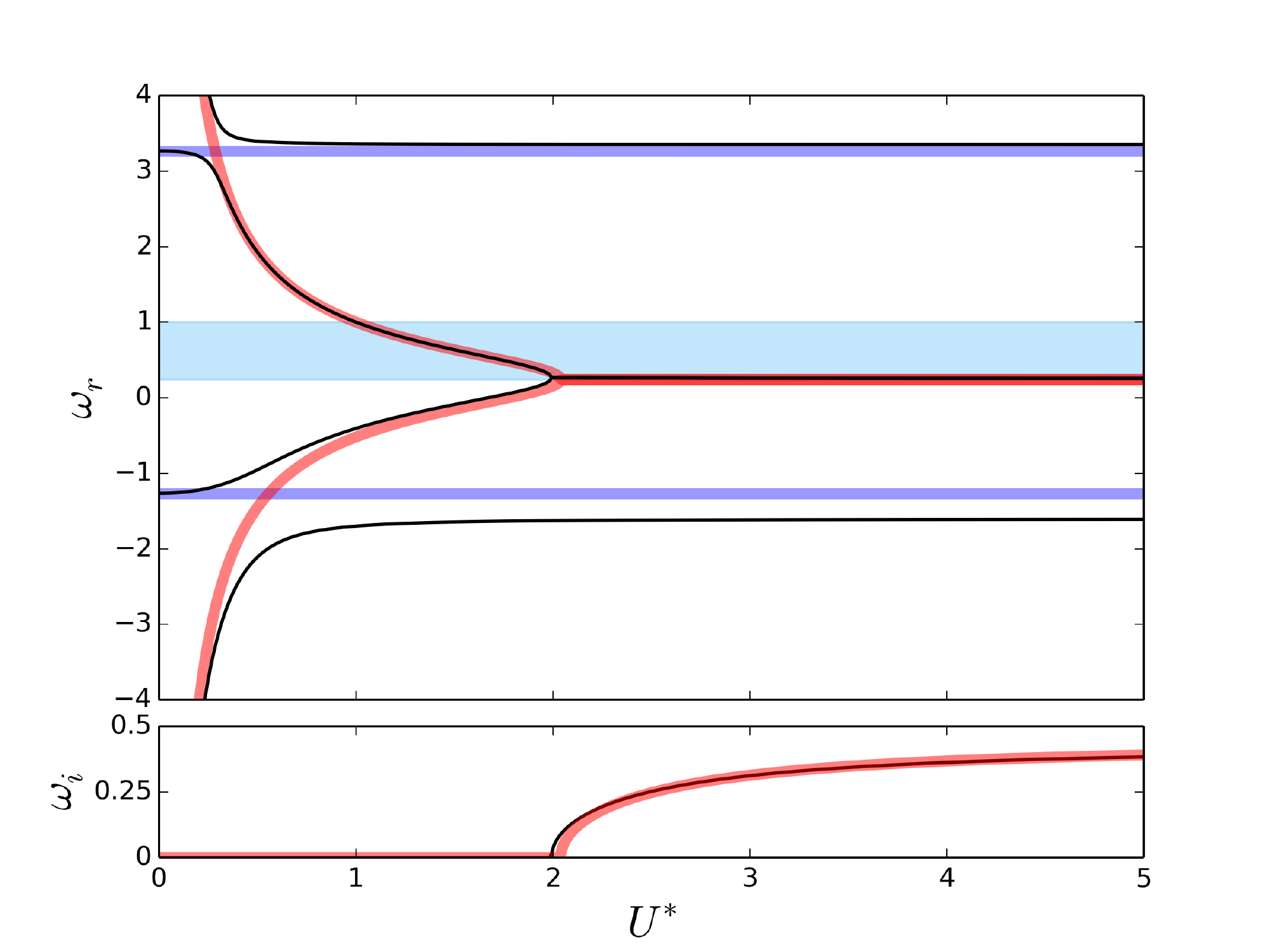} &
\includegraphics[width=6cm, trim = 0cm 0cm 1.8cm 0cm, clip]{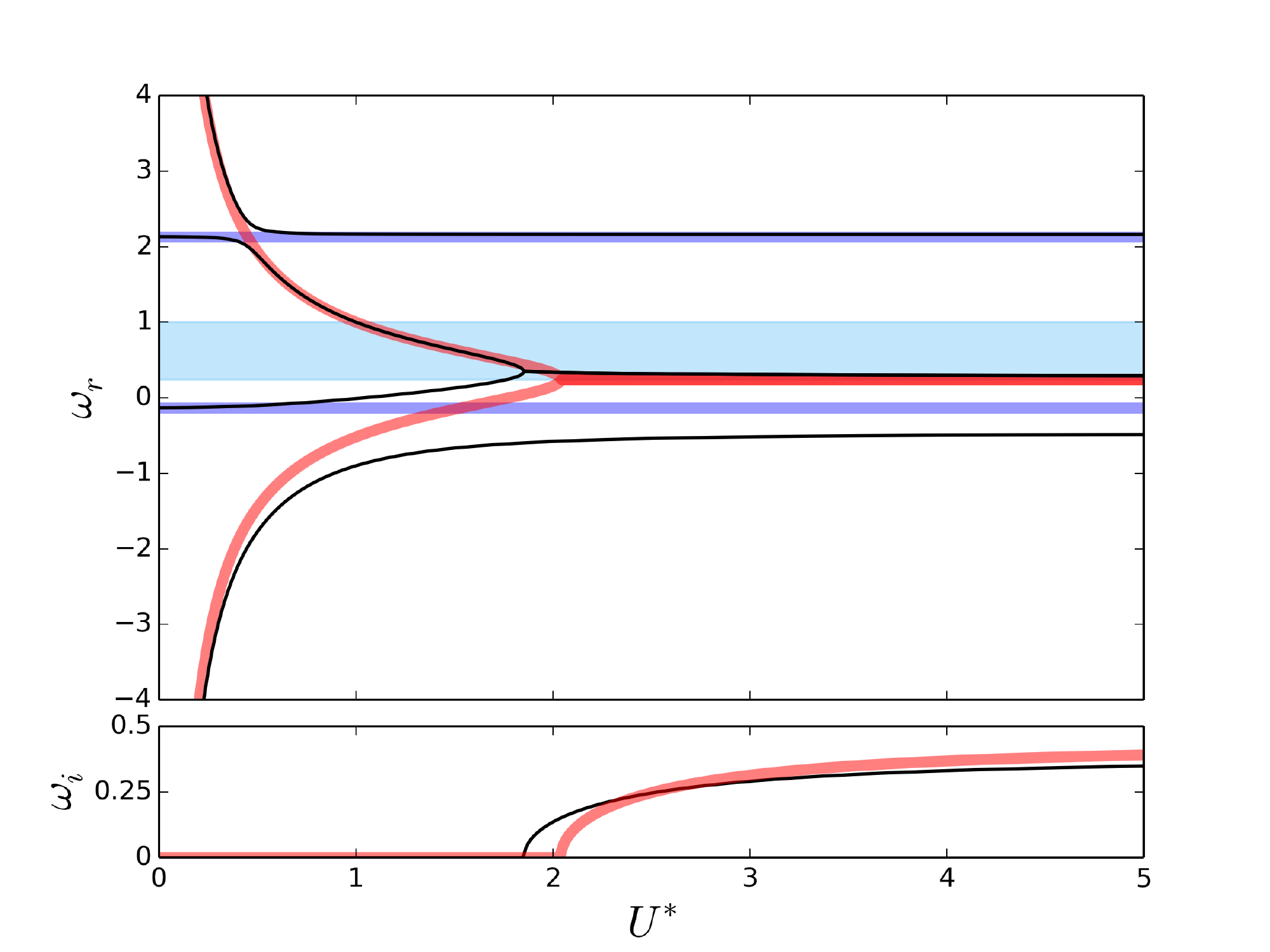} \\
$(a)$  $\mbox{Fr}=0.3$ & $(b)$ $\mbox{Fr}=0.6$ \\
\includegraphics[width=6cm, trim = 0cm 0cm 1.8cm 0cm, clip]{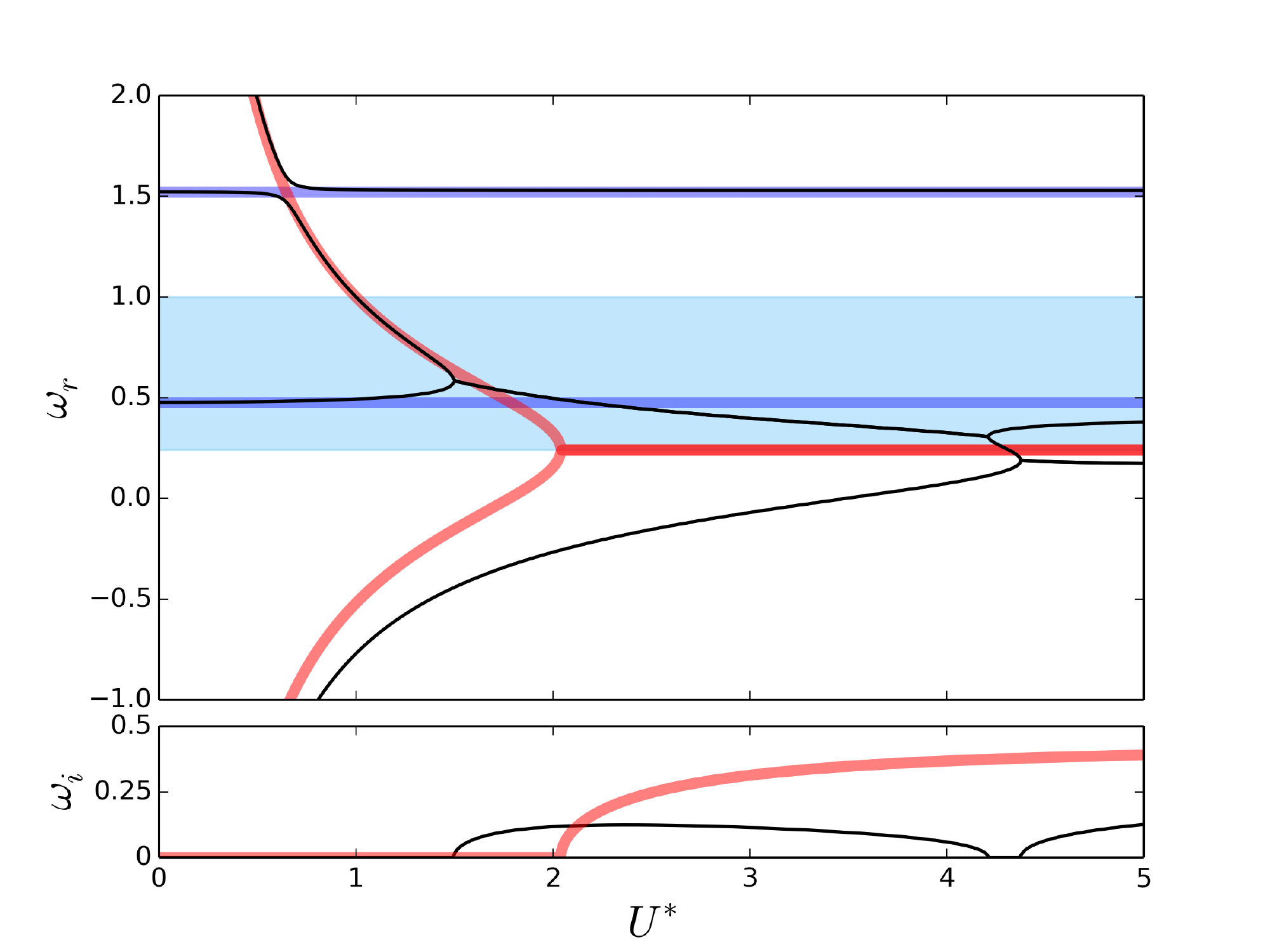} &
\includegraphics[width=6cm, trim = 0cm 0cm 1.8cm 0cm, clip]{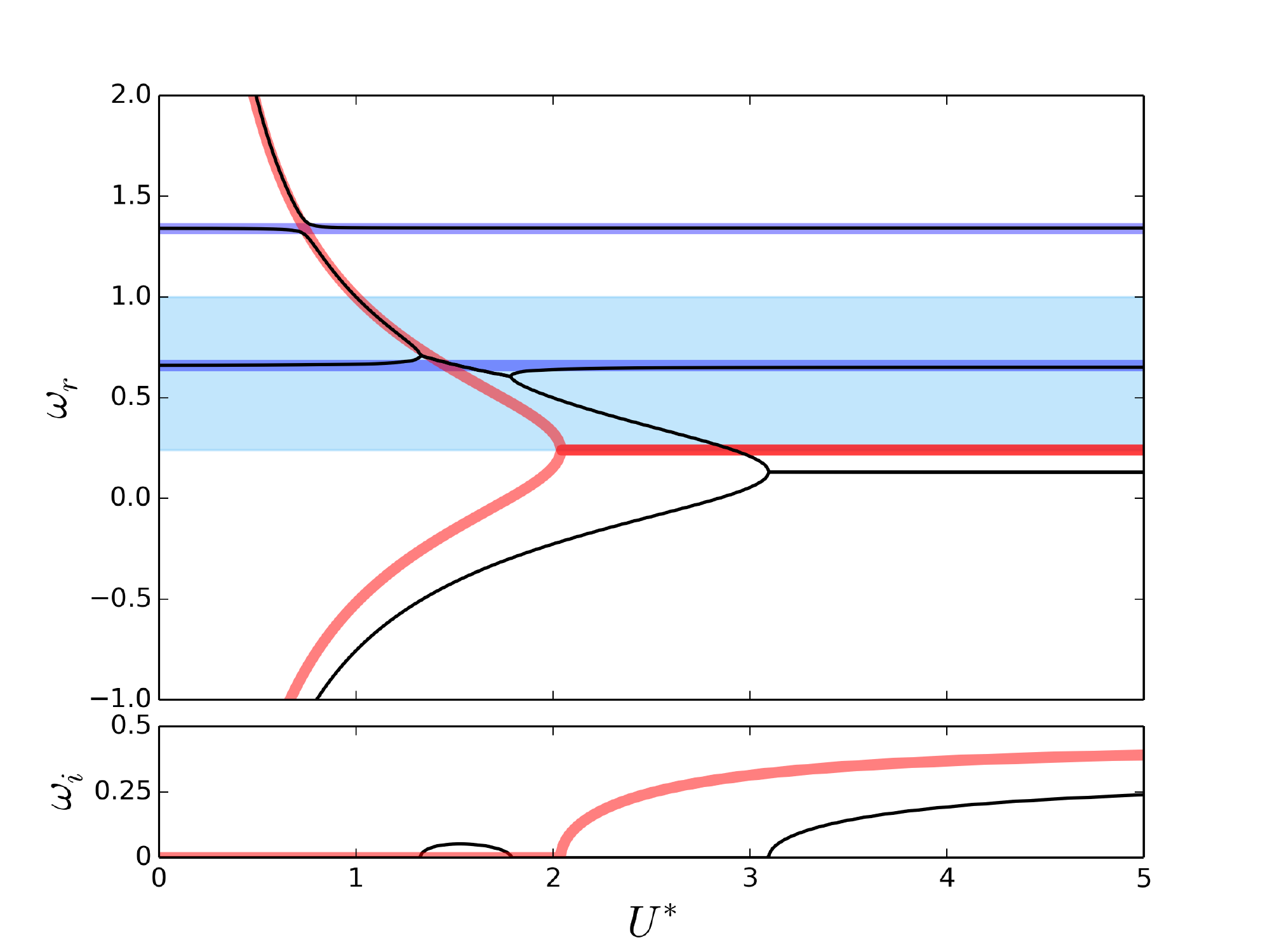} \\
$(c)$  $\mbox{Fr}=1.3$ & $(d)$ $\mbox{Fr}=2$ \\
\includegraphics[width=6cm, trim = 0cm 0cm 1.8cm 0cm, clip]{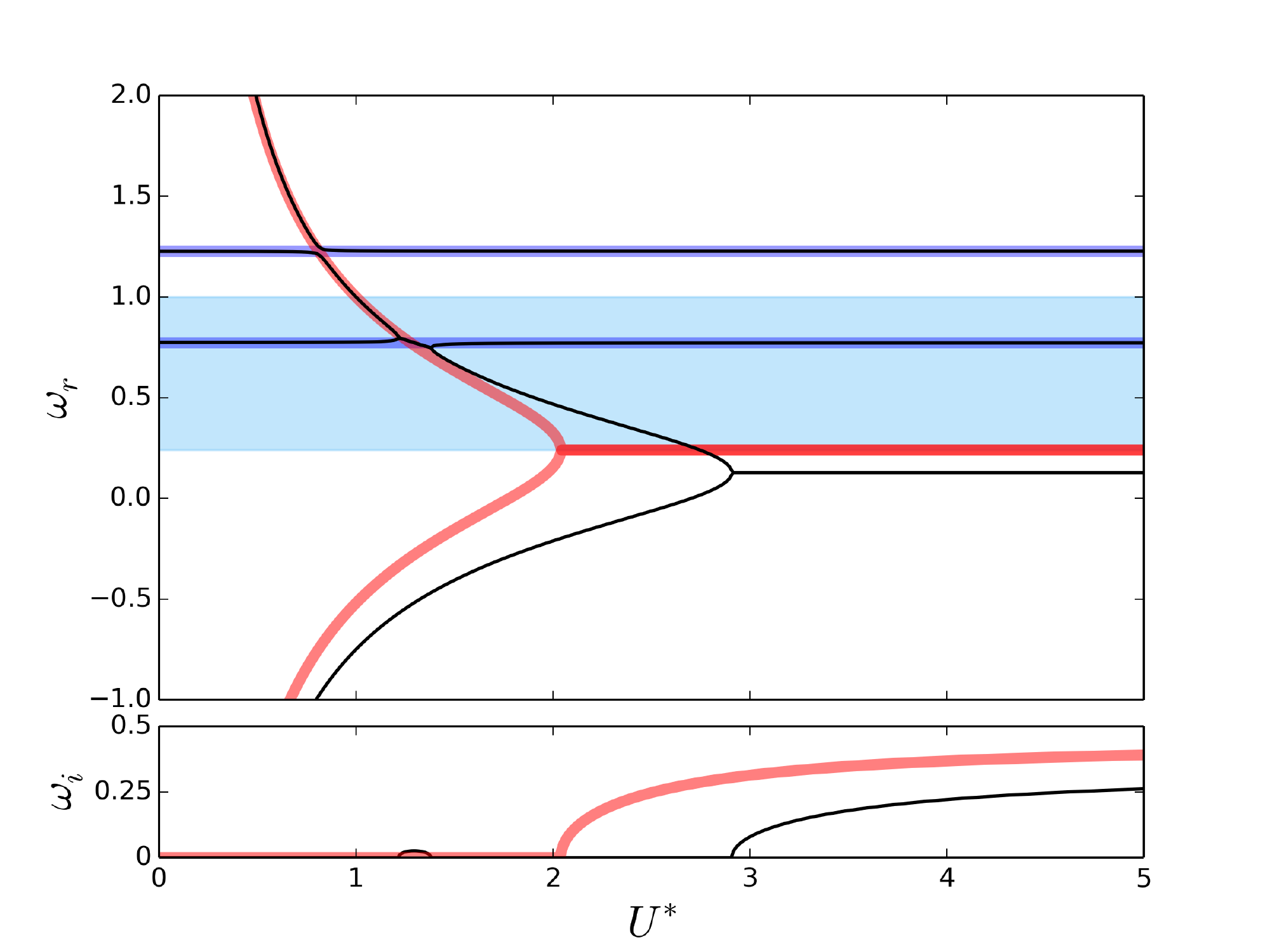} &
\includegraphics[width=6cm, trim = 0cm 0cm 1.8cm 0cm, clip]{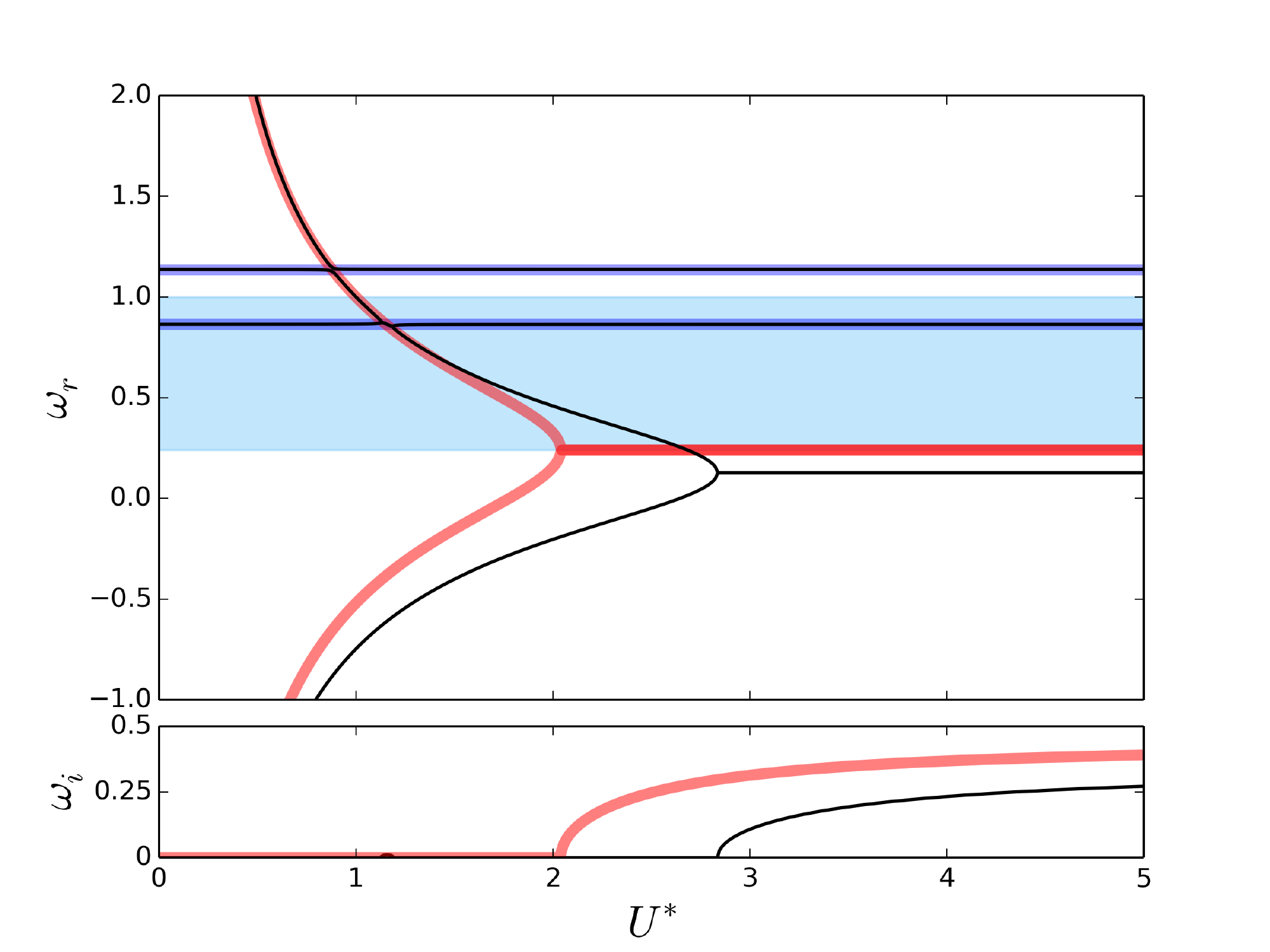} \\
$(e)$  $\mbox{Fr}=3$ & $(f)$ $\mbox{Fr}=5$ \\
\end{tabular}
\caption{\label{fig:freq_Fr} Evolution with $U^*$ of the frequencies of the fully-coupled waves (black, Eq.~\eqref{eq:DfDg}) for various Froude numbers for $h^*=0.5$. The uncoupled flag waves (red, Eq.~\eqref{eq:Df}) and uncoupled gravity waves (blue, Eq.~\eqref{eq:Dg}) are also plotted for comparison. The shaded blue area correspond to the frequency range given by Eq.~\eqref{eq:instab_cond2}.}
\end{figure}

\subsection{Fully-coupled results: frequencies and growth rates}

 We now turn to the fully-coupled problem for which the evolution with $U^*$ of the frequencies and growth rates are reported on Figure \ref{fig:freq_Fr} for several values of $\mbox{Fr}$.  The dispersion relation, Eq.~\eqref{eq:DfDg}, is a quartic equation with four solutions, two of which roughly behaving as uncoupled free surface waves, while the  two others roughly behave as uncoupled flag waves. 

For $\mbox{Fr}=0.3$, Figure \ref{fig:freq_Fr}$(a)$ indeed shows that two solutions of the fully-coupled system match uncoupled flag waves except for frequencies in the vicinity of uncoupled gravity waves, i.e. around $(U^*= 0.3,\, \omega_r= 3.3)$ and $(U^*= 0.5, \,\omega_r= -1.3)$. However, no instabilities appear around crossings between uncoupled solutions: in the coupled case, the two branches seem to merge but do not; in fact, they repel while exchanging identity. This is consistent with the observation that for $\mbox{Fr}=0.3$, these crossings occur outside the unstable frequency range established by Eq.~\eqref{eq:instab_cond2}. The instability threshold for $U^* \approx 2$ is well captured by the uncoupled flag solution and is therefore identified with confined-flag flutter. In that case the free surface essentially behaves as a rigid wall. Similar observations can be made for $\mbox{Fr}=0.6$; additionally, the instability threshold appears to be slightly decreased by the coupling with the free surface. 

For larger values of $\mbox{Fr}$ (Figures \ref{fig:freq_Fr}c--f), results are qualitatively different and two distinct instabilities associated with different ranges of reduced velocities are observed. The first (for small values of $U^*$) results from an interaction between uncoupled flag and free surface waves. For those values of $\mbox{Fr}$,  $\omega_g^-$  belongs to the unstable frequency region predicted in Eq.~\eqref{eq:instab_cond2}, and an instability is found near the frequency crossing with an uncoupled flag wave, confirming the scenario established in Section~\ref{subsec:weak_coupling}. The velocity range corresponding to this instability, as well as the instability growth rate is found to decrease with $\mbox{Fr}$, and this instability is hardly visible for $\mbox{Fr}=5$ (Figure~\ref{fig:freq_Fr}f). This can be understood from inspection of Eq.~\eqref{eq:Dg}: $\omega_g^-\rightarrow 1$ for large $\mbox{Fr}$, which precisely corresponds to the upper bound of the unstable range given by Eq.~\eqref{eq:instab_cond2}.

The second instability (for larger values of $U^*$) is the flutter instability resulting from the interaction of two flag waves. The flutter threshold is however significantly larger than that for a rigidly-confined flag, a feature which is attributed in the following to passive deformations of the free surface.  Depending on the value of $\mbox{Fr}$, one or three critical velocities (noted $U_{c1}^*$, $U_{c2}^*$ and $U_{c3}^*$ by increasing magnitude) can thus be identified as the limits between stable and unstable regimes (Figure~\ref{fig:freq_Fr}). 
\begin{figure}
\centering
\begin{tabular}{cc}
\includegraphics[width=6cm, trim = 1.3cm 0cm 2.2cm 0cm, clip]{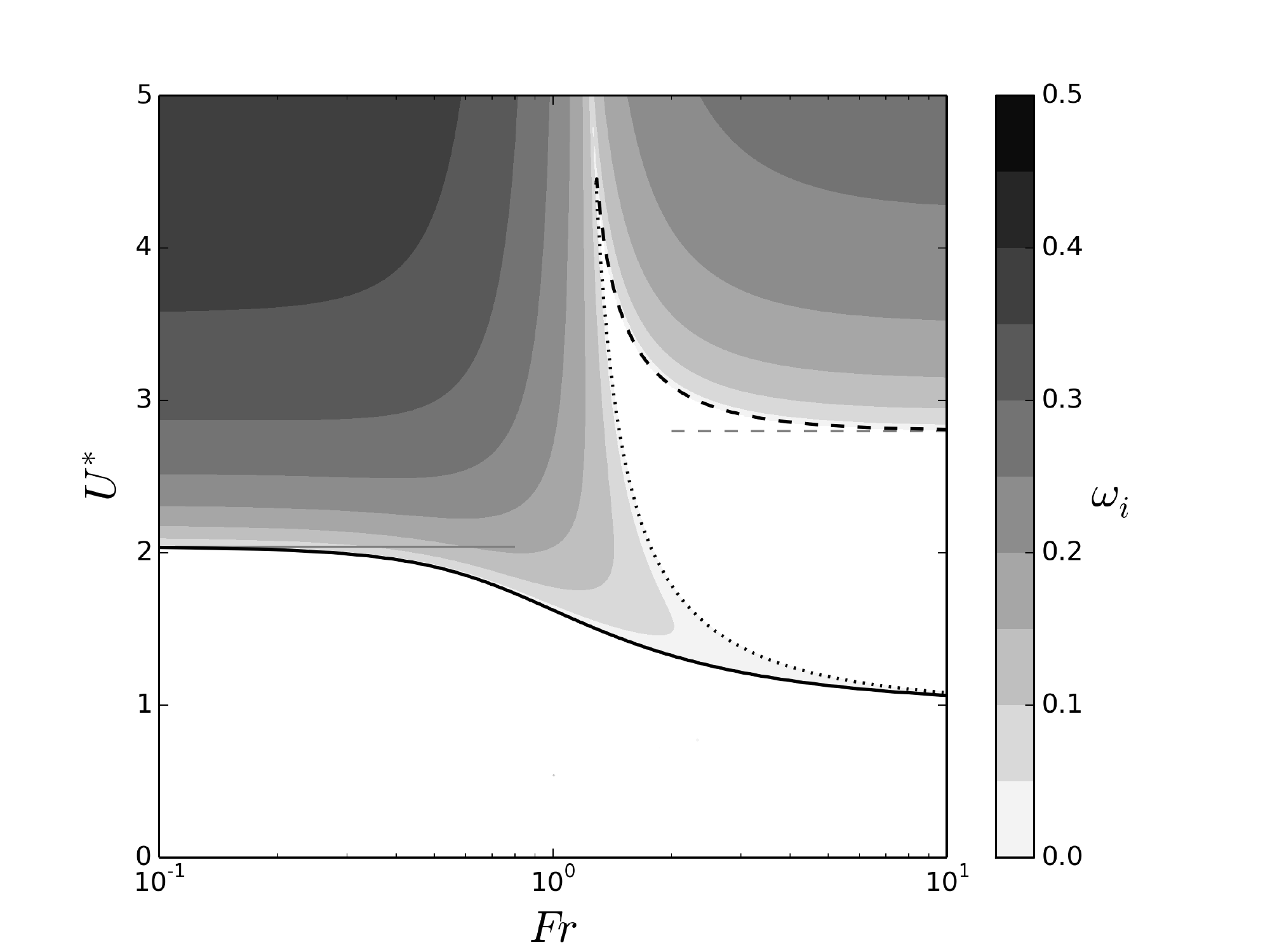} &
\includegraphics[width=6cm, trim = 1.3cm 0cm 2.2cm 0cm, clip]{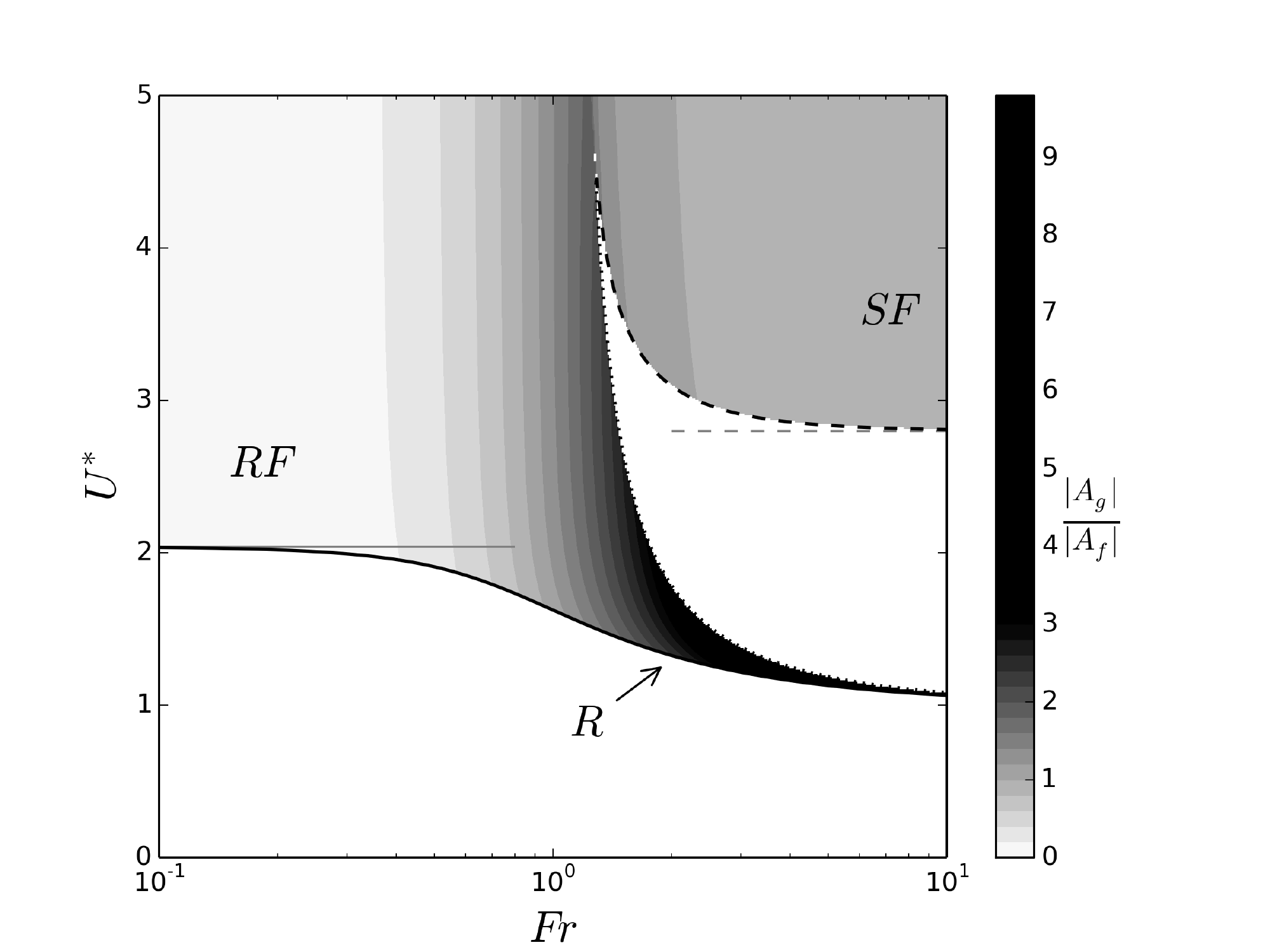} \\
$(a)$ & $(b)$
\end{tabular}
\caption{\label{fig:maps_Fr} Evolution of the growth rate of the unstable coupled mode $(a)$ and corresponding amplitude ratio $(b)$ with $\mbox{Fr}$ and $U^*$ for $h^*=0.5$. Black lines correspond to velocity thresholds $U_{c1}^*$ (plain),  $U_{c2}^*$ (dotted) and  $U_{c3}^*$ (dashed). {Thin horizontal lines indicate flutter thresholds in the small-$\mbox{Fr}$ (plain, $U_{c}^*=2.04$) and large-$\mbox{Fr}$ (dashed,  $U_{c}^*=2.80$) limits.}}
\end{figure}

\subsection{Fully-coupled results: unstable modes}
The unstable regimes identified above can be characterized further by the relative amplitude $|A_g/A_f|$ and phase shift $ \phi_{gf}$ between free surface and flag displacements:
\begin{equation}
\label{eq:AgsurAf}
\frac{A_g}{A_f} = \left|\frac{A_g}{A_f} \right| e^{i \phi_{gf}} = \frac{\mbox{Fr}^2 (\omega -1)^2}{\sinh h^* \left[\mbox{Fr}^2 (\omega-1)^2 \coth h^* - 1 \right]} \cdot
\end{equation} 
Contours of growth rate and amplitude ratio are reported on Figure \ref{fig:maps_Fr} for varying $\mbox{Fr}$ and $U^*$, and three different unstable regimes can be identified.\\ 

\noindent \textit{Rigidly-confined flutter ${(RF)}$.} 

\begin{figure}
\centering
\begin{tabular}{ccc}
\includegraphics[width=3.5cm, trim = 0cm 0cm 0cm 0cm, clip]{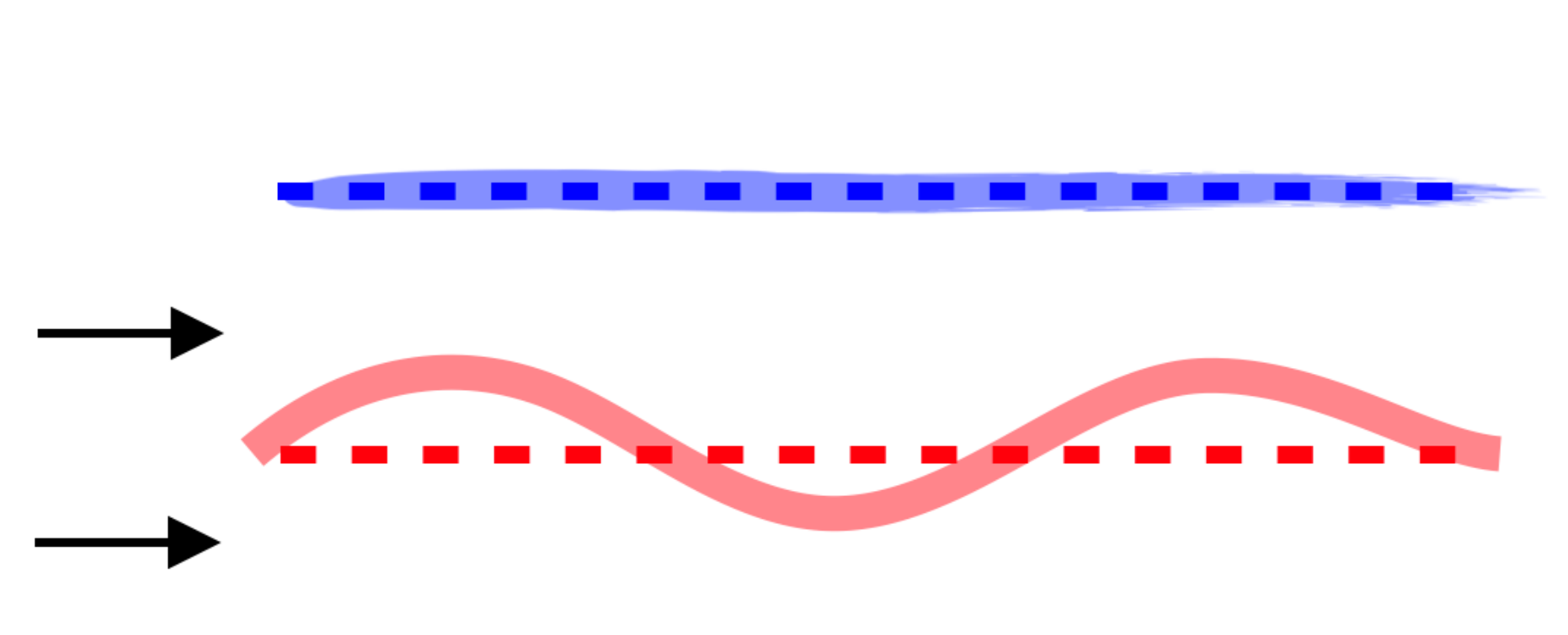} &
\includegraphics[width=3.5cm, trim = 0cm 0cm 0cm 0cm, clip]{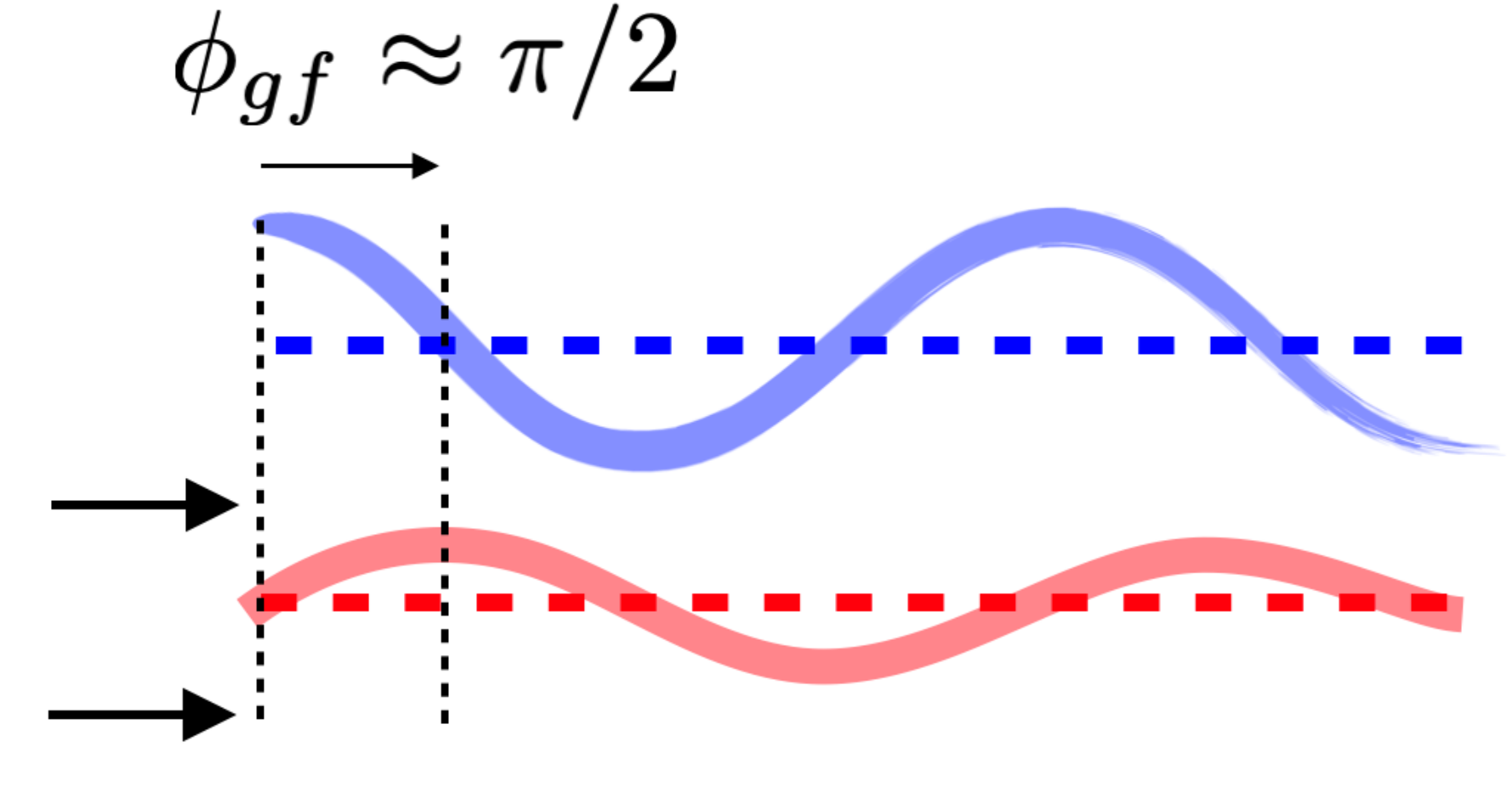} &
\includegraphics[width=3.5cm, trim = 0cm 0cm 0cm 0cm, clip]{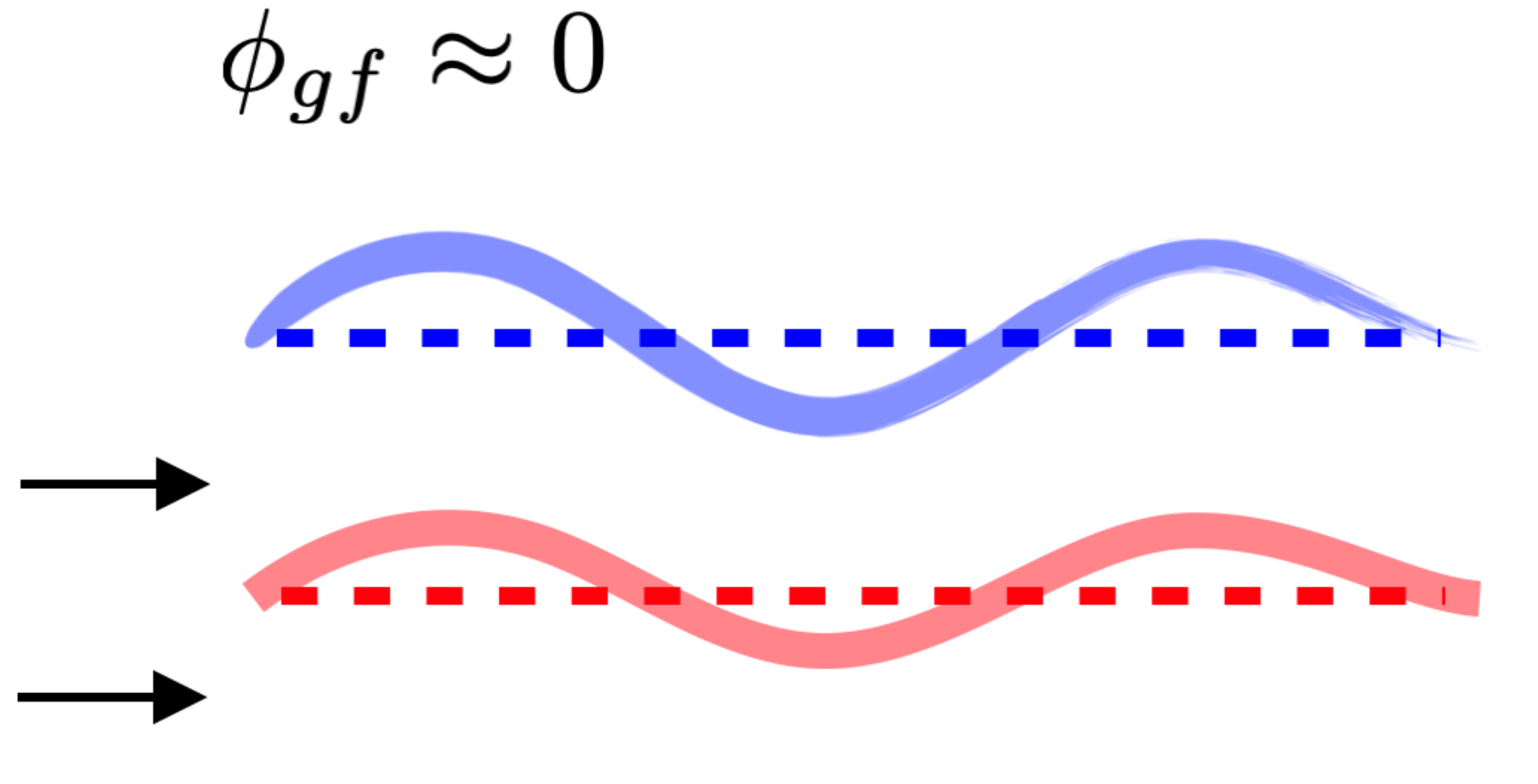} \\
$(a)$Regime $RF$ & $(b)$ Regime $R$ & $(c)$ Regime $SF$
\end{tabular}
\caption{\label{fig:unstable_regimes} Sketches of the unstable regimes: $(a)$ {Rigidly-confined flutter}, $(b)$ {Flag/free surface resonance}, $(c)$ {Softly-confined flutter}.}
\end{figure}

Unstable modes obtained for small values of $\mbox{Fr}$ have small amplitude ratios, i.e. negligible free surface motion, see Eq.~\eqref{eq:AgsurAf}. These correspond to the classical flutter instability in confinement by a rigid wall and are thus termed \textit{rigidly-confined flutter} or $RF$ (white unstable region on Figure \ref{fig:maps_Fr}$b$, i.e. for  small $\mbox{Fr}$ and $U^*>U_{c1}^*$). More specifically, the instability results from the coupling of two flag waves and the frequency satisfies $\mbox{Fr}^2 (\omega-1)^2 \coth h^* \ll 1$. Under this condition, Eq.~\eqref{eq:flargespan} indeed shows that the uncoupled flag problem described in section~\ref{subsec:uncoupled_solutions} is recovered. 
The free surface then behaves as a rigid wall if the flutter frequency falls inside, and far from, the frequency range $[\omega_g^-, \omega_g^+]$. 
The surface's leading order effect reduces to a classical confinement by a rigid surface (Figure~\ref{fig:unstable_regimes}$a$) which destabilizes the flag when the confinement is increased as a result of the increased added mass (as already discussed in Section~\ref{subsec:uncoupled_solutions}).
{The thin horizontal plain line on Figure \ref{fig:maps_Fr} corresponds to the analytical threshold obtained for $\mbox{Fr}=0$ (Eq.~\eqref{eq:Uc}) i.e. $U_{c}^*=2.04$ for $M^*=0.1$ and $h=0.5$; it clearly shows that the free surface can be considered as rigid for $\mbox{Fr}\lesssim 0.3$ for the present set of parameters.} \\

\noindent{\textit{Flag/free surface resonance ($R$).}}

The second unstable regime, obtained for intermediate $\mbox{Fr}$, results from the coupling of the slow free surface wave with a flag wave of similar frequency, thus corresponding to a \textit{flag/free surface resonance} or regime $R$. The displacement of the free surface is large compared to that of the flag (dark regions on Figure~\ref{fig:maps_Fr}$b$). 
A schematic illustration of regime $R$ is shown on Figure~\ref{fig:unstable_regimes}$(b)$.  
A smooth transition is observed between regime $RF$ and regime $R$ as $\mbox{Fr}$ is increased (Figure~\ref{fig:maps_Fr}). 

This regime is expected for frequencies close to uncoupled surface wave solutions  $D_g(\omega)=0$, i.e. for frequencies such that $A_g/A_f \gg 1$, see Eq.~\eqref{eq:AgsurAf}. Section~\ref{subsec:weak_coupling} further established that crossing between uncoupled flag and free surface waves lead to an instability, if and only if Eq.~\eqref{eq:instab_cond2} is satisfied, which provides the frequency range in which crossing leads to an instability. Interestingly, this unstable frequency range vanishes for large values of $M^*$ suggesting that this kind of instability cannot be observed for light structures (large $M^*$). \\

\noindent\textit{Softly-confined flutter $(SF)$.} 

The third regime is obtained for large values of $\mbox{Fr}$ and $U^*>U_{c3}^*$ (grey area on Figure~\ref{fig:maps_Fr}). In this regime, the amplitude of displacement of the flag and free surface are of the same order (see Figures~\ref{fig:maps_Fr}b and \ref{fig:unstable_regimes}c). 
 As seen from Figure~\ref{fig:freq_Fr}(d--f), this unstable regime results from the interaction of two flag waves, and is thus referred to as \textit{softly-confined flutter} or $SF$. As for regime $RF$, the difference with classical flutter stems from the presence of the free surface at a distance $h^*$ above the flag. 
 
In regime $SF$, the destabilizing interaction of two structural waves occurs at a frequency such that $\mbox{Fr}^2 (\omega-1)^2 \coth h^* \gg 1$, which therefore lies outside the range limited by the free surface frequencies, $[\omega_g^-,\omega_g^+]$. As a result, the amplitude ratio defined in Eq.~\eqref{eq:AgsurAf} is close to $1/\cosh h^*$, and approaches unity when the flag lies just under the free surface, but becomes vanishingly small when the flag's immersion increases. When the flag is located close to the free surface, this results in displacements of the flag and free surface of similar magnitude, which are also in-phase in this regime: the free surface essentially follows the displacement of the flag  (see Figure~\ref{fig:unstable_regimes}c). 
 
In addition, for  $\mbox{Fr}^2 (\omega-1)^2 \coth h^* \gg 1$, the fluid loading becomes $f = m_a/2$ with the added mass $m_a =\tanh h^* + 1$, see Eq.~\eqref{eq:flargespan}. The dynamics is therefore perfectly analogous to regime $RF$, with a velocity threshold of the form $U_{c}^* = \sqrt{1 + [m_a M^*]^{-1} }$, except that the confinement has the opposite effect. {This large-$\mbox{Fr}$ analytical limit is indicated as a thin horizontal dashed-line on Figure~\ref{fig:maps_Fr} ($U_{c}^*=2.80$) which shows that softly-confined flutter is obtained for $\mbox{Fr}\gtrsim 5$.}
 For $h^*\ll1$, $m_a \sim h^* + 1$ with the two contributions in the added mass coming from both sides of the flag. The whole mass of the fluid layer between the flag and the free surface therefore acts in the added mass. This is easily understood from the other characteristics of this regime as the whole upper fluid layer perfectly follows the flag displacements and therefore undergoes the same accelerations. The added mass therefore decreases if the thickness $h^*$ of  upper fluid is decreased, and the single-sided flag regime is smoothly reached for $h^*\rightarrow 0$. 

In regime $SF$, the presence of the free surface stabilizes the flag, which is exactly opposite to its effect in regime $RF$. 
Additionally it is worth mentioning that, as for regime $RF$,  the effect of the confinement is reduced for large values of $M^*$ (light flags).

\section{Finite-span configurations}
\label{sec:finite}
Section~\ref{sec:largespan} provided critical physical insight on the flag-free surface interaction in the limit of infinite span (which results in a two-dimensional problem). In this section, we analyse the influence of the flag's finite span numerically {under the assumption that the flag displacement is invariant along its span} (see Figure \ref{fig:sketch}). For numerical reasons associated with the treatment of the bottom boundary, we also assume finite depth, and an additional non-dimensional parameter $d^*=kd$ is introduced, with $d$ a rigid horizontal bottom boundary. This parameter is fixed to $d^*=2$ in the following, a value corresponding to a trade-off between  small enough numerical domain and limited effects of a bottom boundary.  The equations governing the problem remain unchanged, except for the bottom boundary condition, Eq.~\eqref{eq:infinity}, which is replaced by an impermeability condition $\partial \phi / \partial z = 0$ at $z=-d^*$.

As in Section~\ref{sec:largespan}, two uncoupled problems can be defined, namely uncoupled flag and free surface problems (Figure~\ref{fig:sketch_uncoupled}). The dispersion relation of the fully-coupled problem can be written in the form of Eqs.~\eqref{eq:DfDg}--\eqref{eq:Dg} with generalized definitions of $m_a$, ${\cal F}$ and ${\cal C}$ (see ~\ref{app} for more details), providing a direct generalization of the weak coupling analysis performed for the large-span limit.

Note however that ${\cal F}$ and ${\cal C}$, Eqs.~\eqref{eq:F_finite_size} and \eqref{eq:C_finite_size}, now depend on the structure of the gravity wave solution considered. In particular, the coupling term ${\cal C}$ vanishes exactly for  gravity waves that are anti-symmetric across the channel (\ref{app}). As a result, such modes cannot lead to \textit{flag/free surface resonances} as they do not produce any net forcing on the flag once integrated along the span-wise direction. For symmetric modes, the weak coupling analysis predicts instabilities near crossings of uncoupled solutions, with the additional condition that such crossing occurs in the frequency range given by Eq.~\eqref{eq:instab_cond2} with $m_a$ given by Eq.~\eqref{eq:added_mass_finite_span}. 

\begin{figure}
\centering
\includegraphics[width=12cm, trim = 0cm 0cm 0cm 0cm, clip]{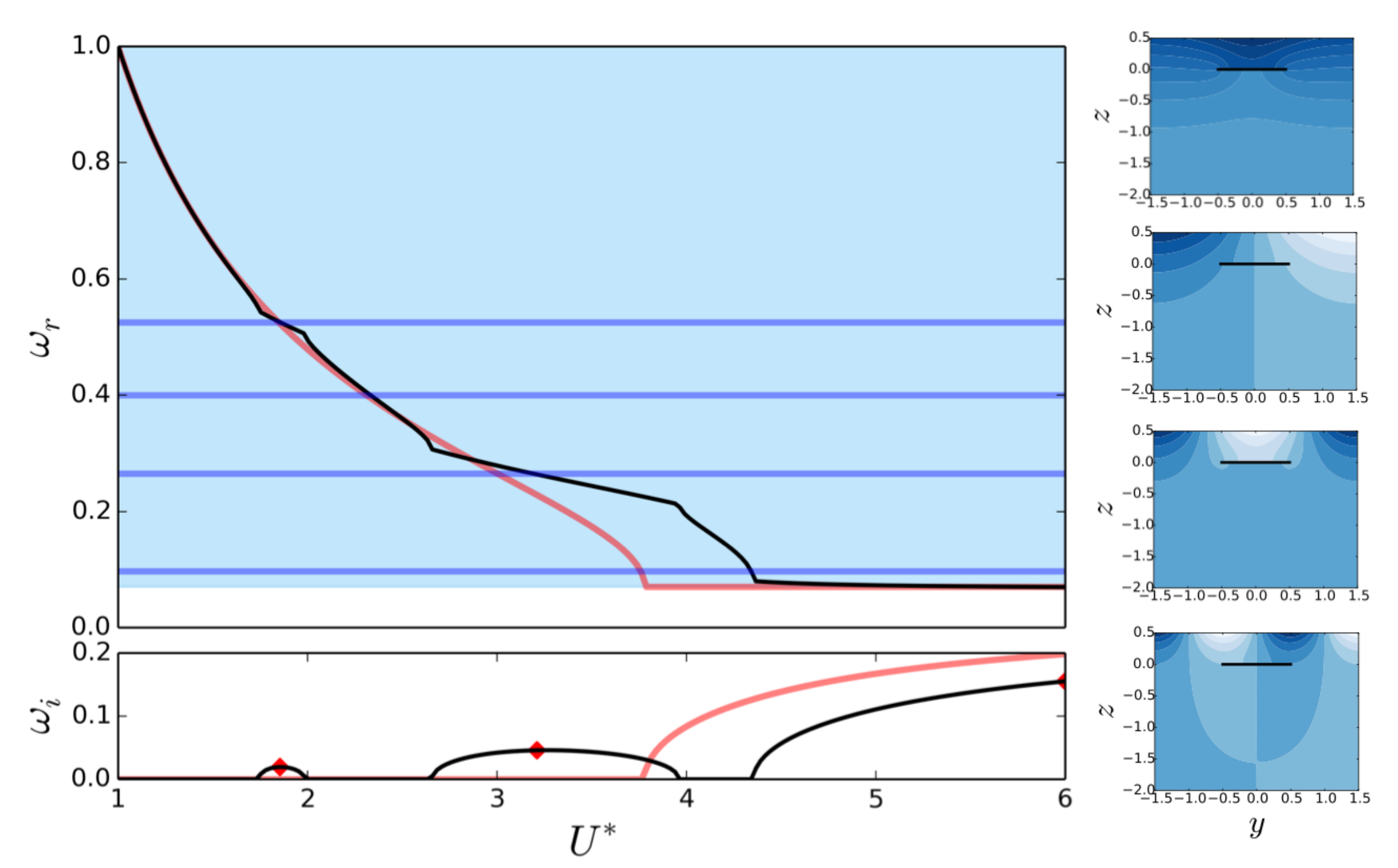}
\caption{\label{fig:freq_finitespan} Evolution with the reduced velocity $U^*$ of the fully-coupled modes for $l=1$, $d=2$ and $\mbox{Fr}=2$. (Left) The top panel shows the frequencies of the fully-coupled problem (black), uncoupled flag waves (red) and uncoupled gravity waves (blue), with the bottom panel indicating the corresponding growthrates. (Right) Evolution of the uncoupled gravity waves structures indicated as blue lines on the left panel. Red symbols indicate solutions investigated on Figure~\ref{fig:structure_Fr2}. Blue shaded area corresponds to frequency range given by Eq.~\eqref{eq:instab_cond2} with added mass obtained from Eq.~\eqref{eq:added_mass_finite_span}. }
\end{figure}

The solutions of the general fully-coupled dispersion relation and their evolution with $U^*$ are compared to the uncoupled solutions for gravity waves and rigidly-confined flag on Figure~\ref{fig:freq_finitespan}. The unstable frequency range leading to \textit{flag/free surface resonances} is also indicated (blue shaded area). Three unstable ranges of reduced velocity can be identified on Figure~\ref{fig:freq_finitespan}. 
The two unstable areas with smallest $U^*$ arise near crossings within the unstable frequency range (blue shaded area) of two uncoupled waves, and thus correspond to flag-free surface resonances. As expected, gravity waves involved in those crossings are symmetric around the midplane of the channel (Figure~\ref{fig:freq_finitespan}, right).  Note that other crossings between uncoupled solutions can be found inside the instability range, but do not lead to any instabilities as they correspond to anti-symmetric modes (Figure \ref{fig:freq_finitespan}, right). \\

The third unstable range (larger $U^*$) on Figure~\ref{fig:freq_finitespan} corresponds to an interaction between two structural waves and is therefore identified as flutter instability. However, the flutter threshold differs from the rigidly-confined case as a result of significant deformations of the free surface, and this regime seems closer to regime $SF$ introduced in the large-span limit.

\begin{figure}
\centering
\begin{tabular}{ccc}
\includegraphics[width=3.8cm, trim = 0cm 0cm 1cm 0cm, clip]{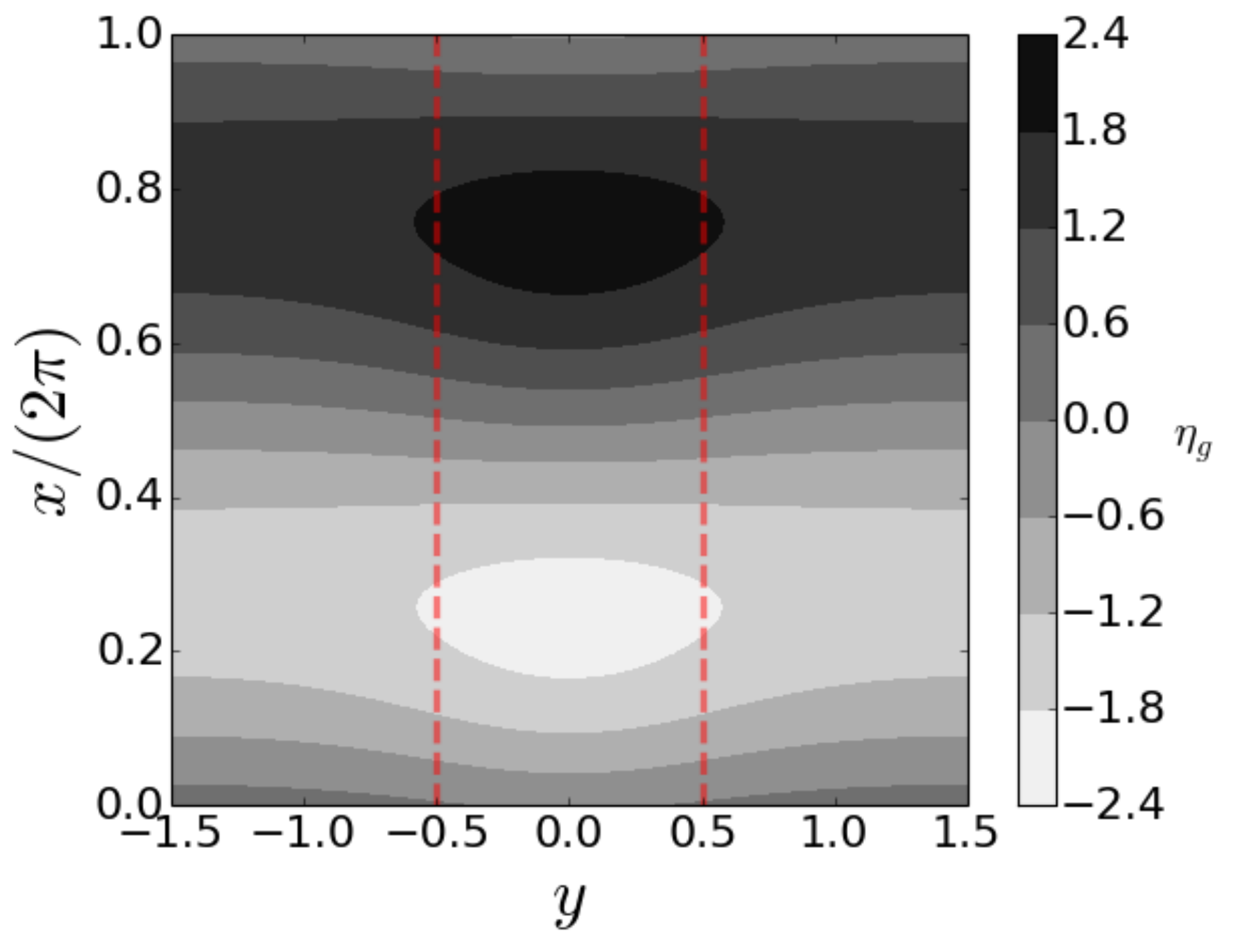} &
\includegraphics[width=3.8cm, trim = 0cm 0cm 1cm 0cm, clip]{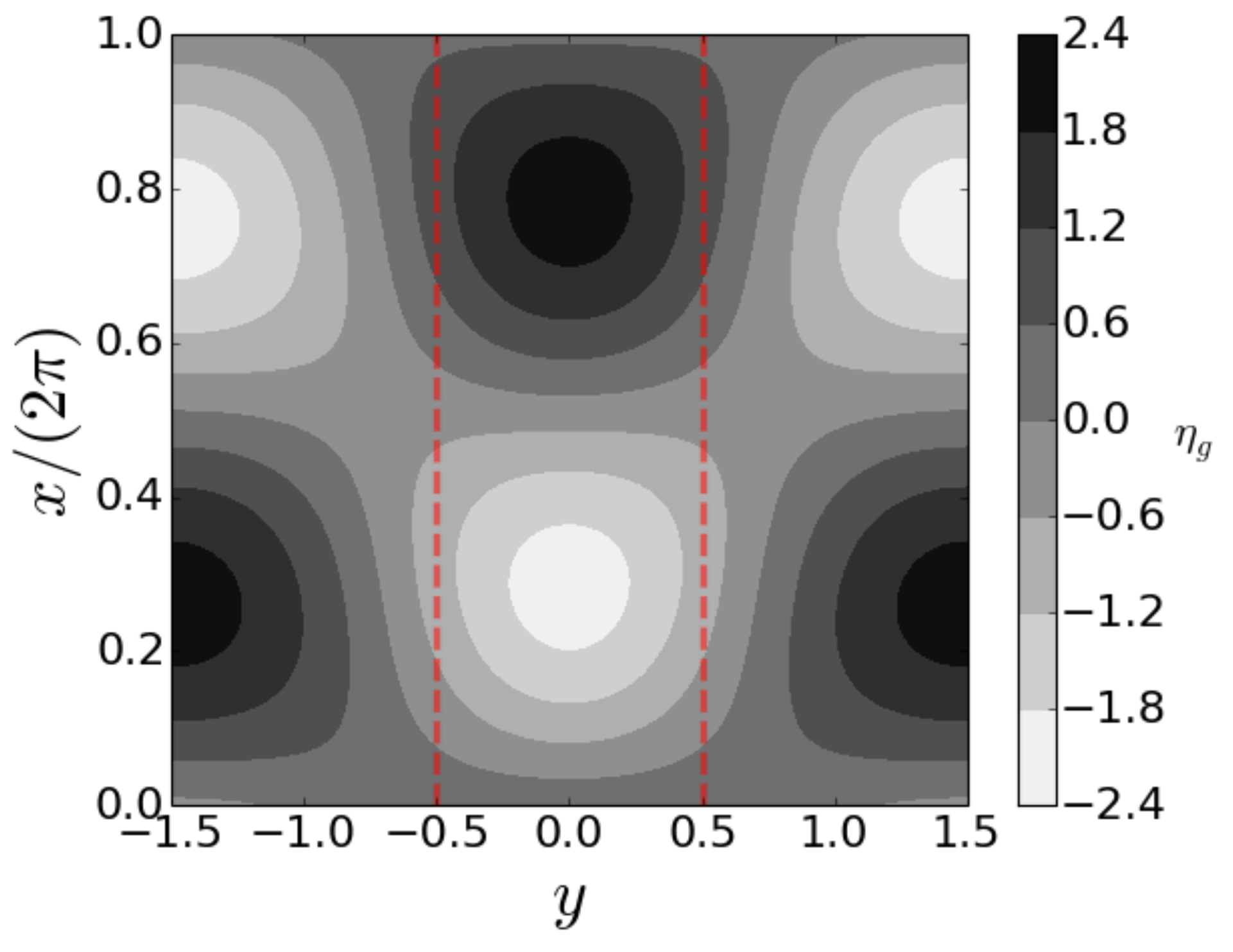}&
\includegraphics[width=3.8cm, trim = 0cm 0cm 1cm 0cm, clip]{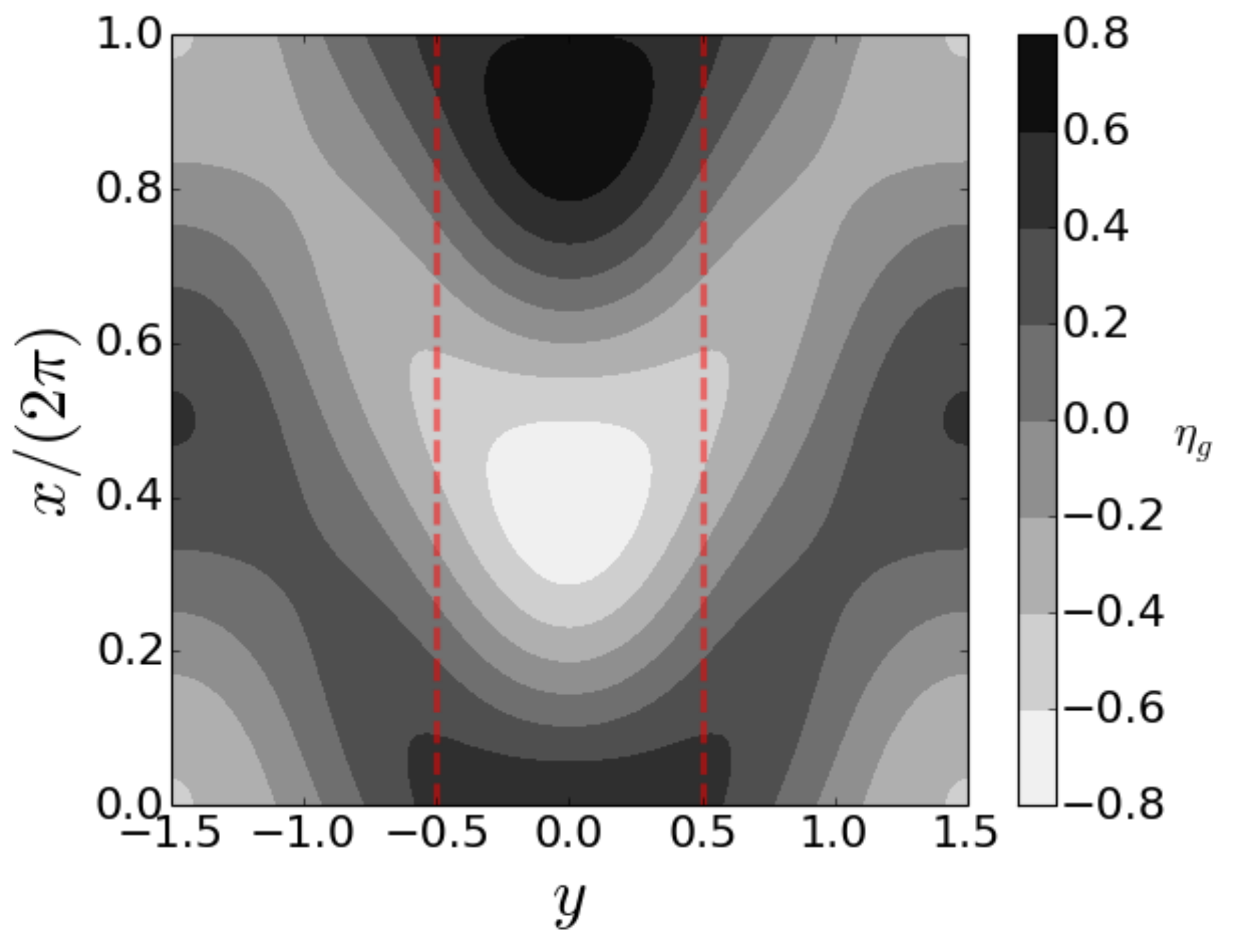}\\
$(a)$ & $(b)$ & $(c)$
\end{tabular}
\caption{\label{fig:structure_Fr2} Free surface deformations of unstable modes for $\mbox{Fr}=2$, $l=1$ corresponding to colored symbols on Figure~\ref{fig:freq_finitespan}. $(a)$ Regime $R_0$, $U^*=1.85$, $(b)$ Regime $R_2$,  $U^*=3.2$ and $(c)$ Regime $SF$, $U^*=6.0$. The flag is located between the two red lines. Flag displacement amplitude is set to unity, and maximum flag displacement corresponds to $x=0$. }
\end{figure}

To support the proposed classification for the different instabilities, free surface deformations of the corresponding unstable patterns are shown on Figure~\ref{fig:structure_Fr2} for ${U^*=1.85,\,\,3.2}$ and $6.0$. The flag is located at a distance $h^*$ below the free surface and in the region $-0.5\leq y\leq 0.5$ (red dashed lines on Figure \ref{fig:freq_finitespan}).
The free surface deformations shown on Figure~\ref{fig:structure_Fr2}$(a)$ and $(b)$ have the same structure as the corresponding symmetric uncoupled free surface waves (Figure~\ref{fig:freq_finitespan}, right). Free surface deformations on Figure~\ref{fig:freq_finitespan} are normalized by the flag displacement amplitude so that the maximum value of the contours can be interpreted as the amplitude ratio between the free surface and the flag (named $A_{gf}$ in this section). Figures~\ref{fig:structure_Fr2}$(a)$ and $(b)$ correspond to relatively large amplitude ratios (respectively about $2$ and $2.4$), in agreement with the classification for \textit{flag/free surface resonances} in the large-span limit (Section~\ref{sec:largespan}).  

In contrast, Figure~\ref{fig:structure_Fr2}$(c)$ shows that free surface deformations for the unstable pattern obtained for large $U^*$ are mainly localized vertically above the flag and with an amplitude ratio of about $0.8$ (Figure~\ref{fig:structure_Fr2}c). This confirms that free surface displacements for that unstable mode are essentially driven by the flag motion and the corresponding instability can indeed be classified in the \textit{softly-confined flutter} regime. 

Some terminology should be introduced to distinguish the different modes of \textit{flag/free surface resonances} that are obtained in the finite-span case. 
As identified for $\mbox{Fr} = 2$ on Figure~\ref{fig:freq_finitespan} and Figure~\ref{fig:structure_Fr2}$(a)$-$(b)$ these instabilities are associated with gravity waves with different span-wise structures. 
More precisely, Figure~\ref{fig:freq_finitespan} (right) shows examples of the corresponding uncoupled gravity waves which can be identified by their number of nodes $(n)$ in the span-wise direction, for instance from $n=0$ to $n=3$ on Figure~\ref{fig:freq_finitespan} (top to bottom). In the following, \textit{flag/free surface resonances} associated to gravity wave with $n$ nodes in the span-wise direction are referred to as $R_n$.

\begin{figure}
\centering
\begin{tabular}{cc}
\includegraphics[width=6cm, trim = 0.5cm 0cm 2cm 0cm, clip]{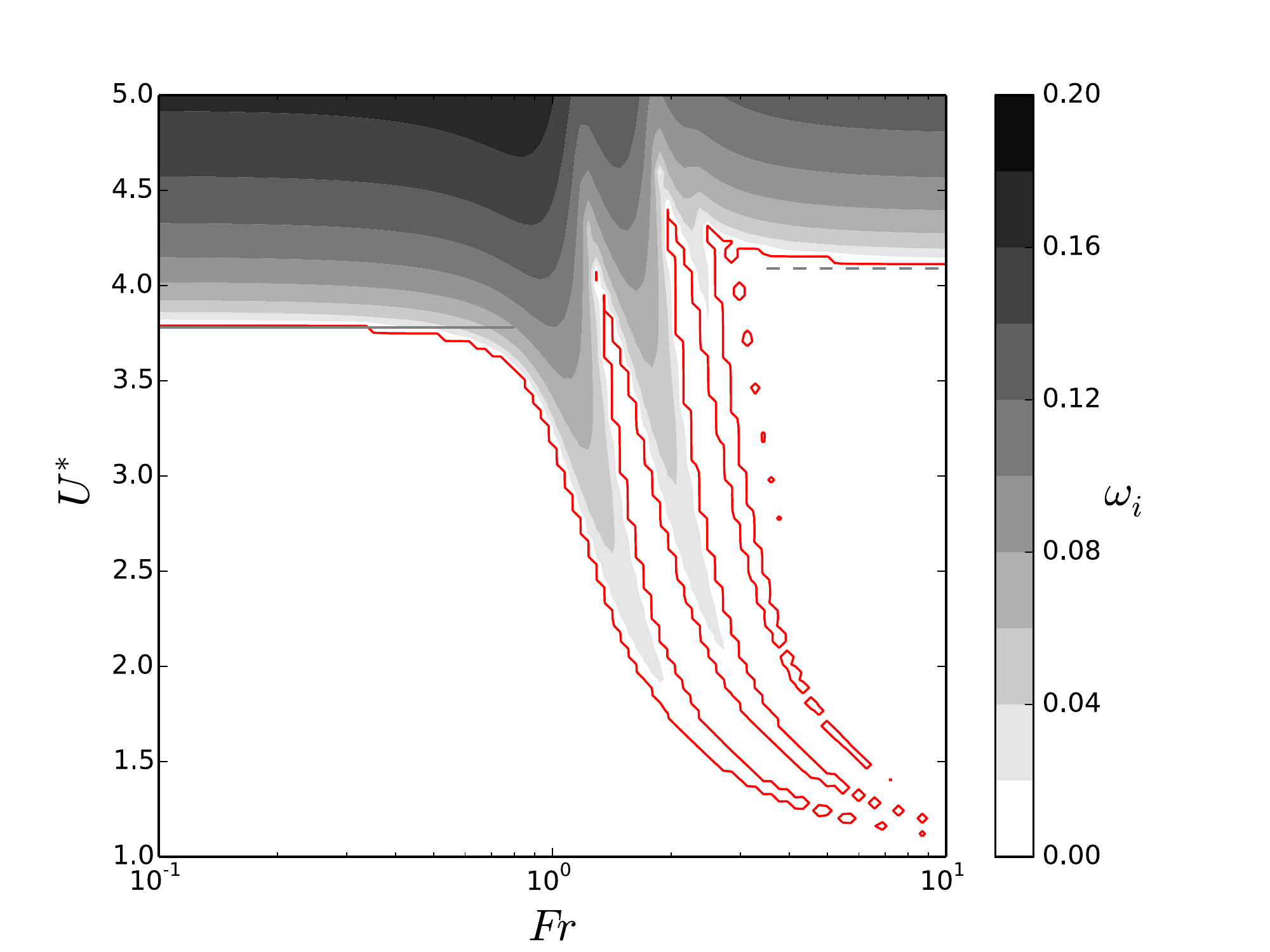} &
\includegraphics[width=6cm, trim = 0.5cm 0cm 2cm 0cm, clip]{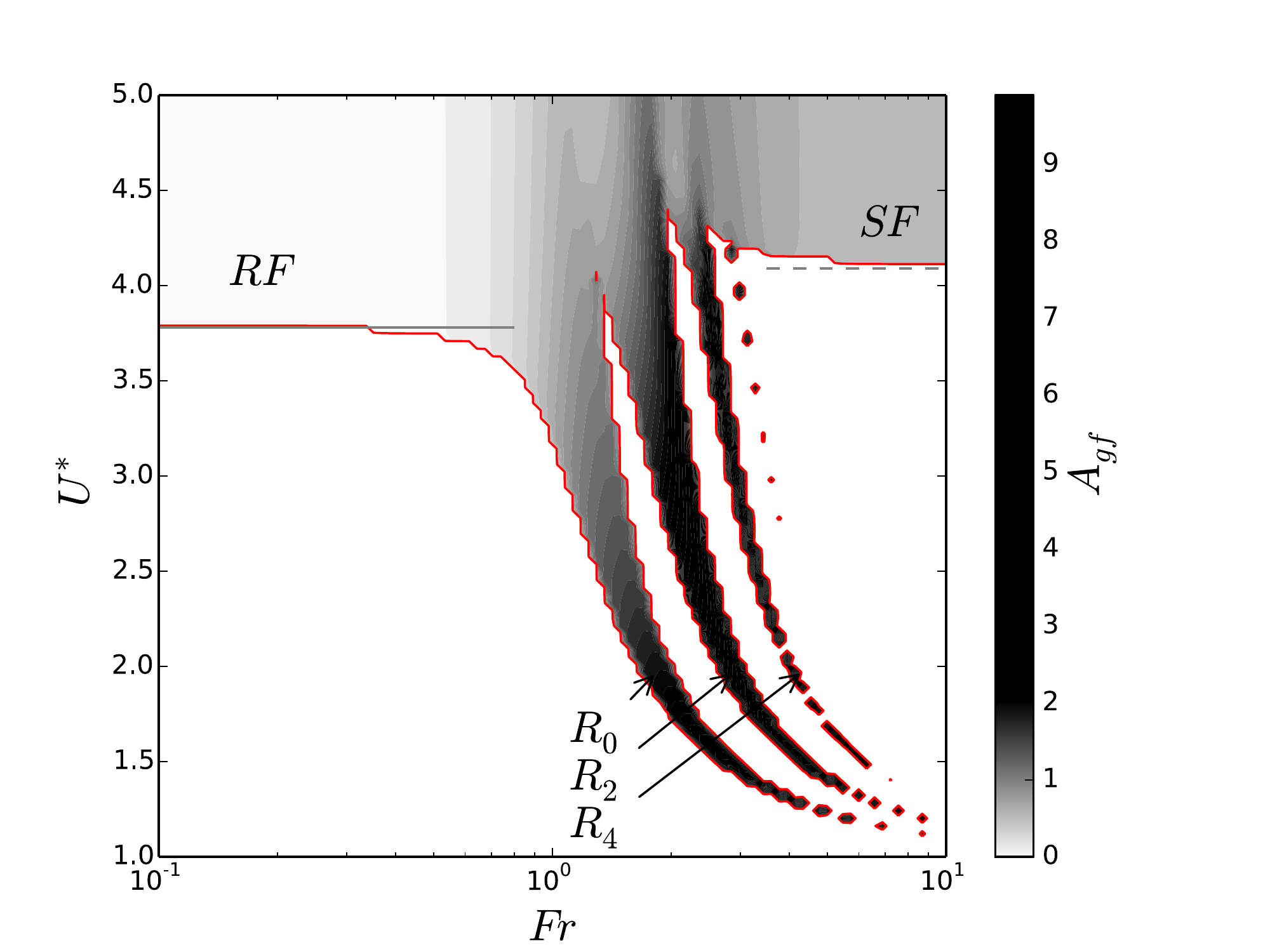}\\
$(a)$ & $(b)$
\end{tabular}
\caption{\label{fig:map_finitespan} Stability maps in the finite span case for $l=1$. $(a)$ Growth rate. $(b)$ Amplitude ratio $A_{gf} = \text{max}[Re(A_g(y) e^{ix}/A_f)]$ as a function of the Froude number $\mbox{Fr}$ and the reduced velocity $U^*$. The stability thresholds are indicated by red lines. {Thin horizontal lines indicate flutter thresholds obtained numerically for $\mbox{Fr}=0$ (plain, $U_{c}^*=3.78$) and $\mbox{Fr}=10^3$ (dashed, $U_{c}^*=4.09$).}}
\end{figure}

The evolution  with $\mbox{Fr}$ and $U^*$ of the distinct unstable modes can be followed using maps of the growth rate and amplitude ratios in that parameter space (Figure~\ref{fig:map_finitespan}). The qualitative similarity with large-span results of Figure~\ref{fig:maps_Fr} should first be noted. In particular, using the amplitude ratio map, the three type of unstable regimes (rigidly confined flutter, softly confined flutter and flag/free surface resonances) can be identified as white, grey and black areas, respectively.
 In this finite span case for $l^*=1$, three unstable branches $R_0$, $R_2$ and $R_4$ can be tracked in the  $\mbox{Fr}$--$U^*$ parameter space  using Figure~\ref{fig:map_finitespan}. 
Modes with an  odd number of nodes are anti-symmetric and do not lead to instabilities. 

An additional effect of the finite span visible on Figure~\ref{fig:map_finitespan} corresponds to the increase of flutter thresholds corresponding to both $RF$ (small $\mbox{Fr}$) and $SF$ (large $\mbox{Fr}$) compared to the large-span case (Figure~\ref{fig:maps_Fr}). This trend is in qualitative agreement with the effect of aspect ratio on classical flutter thresholds for an unbounded flag~\cite{eloy2007}.


\section{Conclusion}
\label{sec:conclusions}

This work analysed the effects of the proximity of a free surface (which can thus carry gravity waves) on the stability of a flexible plate placed in a steady current using local stability analysis.  {The simplified large-span limit, which results in a two-dimensional problem, was first considered before generalizing the discussion to finite-span flags (but with neglected span-wise deformation) placed in a finite-width free surface channel.} The effect on the stability of the Froude number (i.e. the ratio of the incoming flow speed to the characteristic velocity of gravity waves) as well as that of the reduced velocity were investigated, and we identified three main unstable regimes as the Froude number was increased. 

In the first regime, obtained for low $\mbox{Fr}$ (i.e. strong gravitational effects), the free surface behaves essentially as a rigid wall; this \textit{rigidly-confined flutter} corresponds to the flutter instability of a flag within lateral confinement as investigated by \cite{guo2000, doare2012, alben2015, dessi2015}. 
The second unstable regime, \textit{flag/free surface resonances}, is mainly obtained for intermediate Froude numbers (i.e. when surface waves and flow speed are comparable), and results from resonances between surface waves and structural bending waves (both modified by the uniform flow), and is generally associated to large free surface deformations compared to that of the flag. Only one such resonance occurs in the large-span case, but a discrete set of such instabilities is obtained for a {finite-span flag in a finite-width free surface channel}, due to resonances involving free surface waves with more complex span-wise structures. 
The third regime, \textit{softly-confined flutter}, obtained for large $\mbox{Fr}$ (i.e. small gravitational effects) also corresponds to the flag flutter, but in contrast with the first regime, the free surface may undergo large deformations driven by the flag motion (up to that of the flag for small immersion depth). 

The main difference between these regimes is found in the behaviour of the free surface: while it plays a purely passive role in the rigidly-confined and softly-confined flutter (where it is either fixed and flat or entirely driven by the flag motion), it takes an active role when the frequency of the instability lies close to a fundamental frequency of the surface waves. Interestingly, the passive behaviour of the free surface may result in significantly different impact on the flag's stability, depending on the regime and $\mbox{Fr}$: \textit{softly-confined flutter} corresponds to a  stabilizing effect of the free surface (compared to the flag in unbounded flow) while \textit{rigidly-confined flutter} is destabilizing as already reported in the literature in the case of rigid walls~\cite{guo2000, doare2012, alben2015, dessi2015}. As a result, the presence of a coupling with a free surface may either decrease or increase flutter thresholds, or lead to new instabilities. 


From the energy harvesting view point, only flutter regimes (both \textit{rigidly-confined flutter}  and \textit{softly-confined flutter}) appear to be of interest as flag/free surface resonances are associated to small flag displacements and therefore to a small energy transfer from the fluid to the solid. 
Recalling that working conditions of flag-like energy harvesters are closely related to velocity thresholds, the  conclusions of the present work show that the proximity of a free surface may either enlarge (small $\mbox{Fr}$) or reduce (large $\mbox{Fr}$) the operating range of such devices. 
Additionally, the present analysis which combines a flexible structure, a free surface and an incoming flow, could be used to investigate the effects of an incoming flow on wave energy harvesters such as~\cite{alam2012, lehmann2013}, and determine whether such \textit{carpet of wave energy recuperation} could be used as a hybrid wave/current energy harvesting device.


Last but not least, the unstable regimes identified in this work have only been studied experimentally in the low Froude limit consisting of a rigid wall~\cite{dessi2015}.
Such experiments necessarily include finite-length effects that are likely critical both from a fundamental and application points of view and deserve to be investigated further. Finite flag length effects indeed promote complex flow features ranging from hydraulic jumps, wave radiation or wake/free surface interaction, which are not accounted for in the present analysis and may yet play an important role in the dynamics of the system. Modeling and analyzing  these effects represent a major challenge for future research. 


\section*{Acknowledgments}
This work was supported by the European Research Council (ERC) under the European Union's Horizon 2020 research and innovation program (Grant Agreement No. 714027 to S.M.).

\appendix 
\section{Derivation of the dispersion relation for finite span and finite depth}
\label{app}
Details are provided here on the derivation of the dispersion relation under the form \eqref{eq:DfDg} in the finite-span and finite depth case. 
The channel cross section is denoted $\Omega$ with $\mathbf{n}$ the unit vector associated with the base state pointing out of the flow domain on each of the boundaries of $\Omega$, namely the flag surface ($\partial \Omega_f$), the free surface ($\partial \Omega_g$) and the side and/or bottom walls ($\partial \Omega_w$). 
Starting from Eqs.~\eqref{eq:laplace_phir}--\eqref{eq:spanwall_phir}, the perturbation velocity potential is decomposed as $\phi_r = \phi_f + \phi_g$ where $\phi_f$ corresponds to a rigidly confined flag. As a result, $\phi_f$ and $\phi_g$ satisfy:
\begin{align}
\label{eq:laplace_phif}
\frac{1}{{l^*}^2}\frac{\partial^2 \phi_f}{\partial y^2}+ \frac{\partial^2 \phi_f}{\partial z^2} &= \phi_f, \ \ \   \text{on} \ \Omega, \\
\label{eq:freesurface_phif}
\frac{\partial \phi_f }{\partial z} &=0, \ \ \ \ \   \text{on} \  \partial \Omega_g, \\
\label{eq:flag_phig}
\frac{\partial \phi_f}{\partial n} &=  n_z,  \ \ \ \   \text{on} \ \partial \Omega_f,\\
\label{eq:bottomwall_phig}
\frac{\partial \phi_f}{\partial n} &= 0, \ \ \ \ \  \text{on} \  \partial \Omega_w, 
\end{align}
and
\begin{align}
\label{eq:laplace_phig}
\frac{1}{{l^*}^2}\frac{\partial^2 \phi_g}{\partial y^2}+ \frac{\partial^2 \phi_g}{\partial z^2} &= \phi_g, \ \ \  \text{on} \ \Omega, \\
\label{eq:freesurface_phig}
\frac{\partial \phi_g }{\partial z} &= \mbox{Fr}^2 (\omega -1)^2 (\phi_g +\phi_f),  \  \text{on} \ \partial \Omega_g,\\
\label{eq:flag_phig}
\frac{\partial \phi_g}{\partial n} &=  0,  \ \ \ \   \text{on} \ \partial \Omega_f,\\
\label{eq:bottomwall_phig}
\frac{\partial \phi_g}{\partial n} &= 0,  \ \ \ \   \text{on} \ \partial \Omega_w,
\end{align}
with $n_z=\mathbf{e}_z\cdot\mathbf{n}$. 
When the flag is held fixed, $\phi_f=0$, and the second system can thus be interpreted as the uncoupled free surface problem. 
When the flag has non-zero displacement, a forcing term involving flag motion in Eq.~\eqref{eq:freesurface_phig} introduces a coupling between both problems.

Using the decomposition above for the velocity potential, the fluid loading, Eq.~\eqref{eq:f}, can be split into two parts, and Eq.~\eqref{eq:D} can be written in the form
\begin{equation}
\label{eq:Df_Cgf}
\omega^2 -{U^*}^{-2} + m_a M^* (\omega-1)^2  = -M^* (\omega-1)^2 {\cal C}_{gf}
\end{equation}
with 
\begin{equation}
\label{eq:added_mass_finite_span}
m_a = \int_{\partial \Omega_f}{ \phi_f n_z dy}\qquad \textrm{and}\qquad
 {\cal C}_{gf}=  \int_{\partial \Omega_f}{\phi_g dy}.
\end{equation}
Equation~\eqref{eq:Df_Cgf} corresponds to the uncoupled flag dispersion relation (left hand side), coupled with the free surface (right-hand side), with the added mass and coupling coefficients defined in Eq.~\eqref{eq:added_mass_finite_span}.  

The weak form of Eqs.\eqref{eq:laplace_phig}--\eqref{eq:bottomwall_phig} is obtained by projecting these equations along test functions $\phi_g^*$:
\begin{equation}
\label{eq:Dg_Cfg}
 -1 +  {\cal F} \ \mbox{Fr}^2 (\omega -1)^2= -\mbox{Fr}^2(\omega-1)^2 {\cal C}_{fg},
\end{equation}
with 
\begin{equation}
\label{eq:F_finite_size}
{\cal F} = \frac{\displaystyle\int_{\partial \Omega_g}{ \phi_g \phi_g^*dy}}{\displaystyle\int_{\Omega}{ \left( \frac{1}{l^2} \frac{\partial \phi_g}{\partial y}\frac{\partial \phi_g^*}{\partial y} + \frac{\partial \phi_g}{\partial z}\frac{\partial \phi_g^*}{\partial z} + \phi_g \phi_g^* \right)dydz}},
\end{equation}
and 
\begin{equation}
\label{eq:C_finite_size}
{\cal C}_{fg} = \frac{\displaystyle\int_{\partial \Omega_f}{ \phi_f \phi_g^* dy}}{\displaystyle\int_{\Omega}{ \left( \frac{1}{l^2} \frac{\partial \phi_g}{\partial y}\frac{\partial \phi_g^*}{\partial y} + \frac{\partial \phi_g}{\partial z}\frac{\partial \phi_g^*}{\partial z} + \phi_g \phi_g^* \right)dydz}}\cdot
\end{equation}
Multiplying Eqs.~\eqref{eq:Df_Cgf} and \eqref{eq:Dg_Cfg} finally provides the dispersion relation of the fully-coupled system in a form similar to Eq.~\eqref{eq:DfDg} for the infinite-span limit but with a new coupling term which now reads
\begin{equation}
\label{eq:coupling_finite_span}
{\cal C}^2 = M^* \mbox{Fr}^2 (\omega-1)^4 {\cal C}_{fg}{\cal C}_{gf}.
\end{equation}
and with $m_a$ and ${\cal F}$ given by \eqref{eq:added_mass_finite_span} and \eqref{eq:F_finite_size} respectively. 

Using Eq.~\eqref{eq:added_mass_finite_span}, anti-symmetric solutions for $\phi_g$ with respect to $y$ axis (i.e. such that $\phi_g(-y)=-\phi_g(y)$) lead to  ${\cal C}_{gf}=0$, and therefore to ${\cal C}=0$. As a result flag/free surface instabilities cannot arise from a resonance  involving an anti-symmetric free surface wave.


\begin{thebibliography}{10}

\bibitem{paidoussis2004}
M.~P.~Paidoussis. 
\newblock Fluid-structure interactions: Slender structures and axial flows.
\newblock {\em Academic Press, London}, volume 2, 2004.

\bibitem{shelley2011}
M.~J. Shelley and J. Zhang.
\newblock Flapping and bending bodies interacting with fluid flows.
\newblock {\em Annual Review of Fluid Mechanics}, 43:449--465, 2011.

\bibitem{eloy2007}
C. Eloy, C. Souilliez, and L. Schouveiler.
\newblock Flutter of a rectangular plate.
\newblock {\em Journal of Fluids and Structures}, 23(6):904--919, 2007.

\bibitem{giacomello2011}
A. Giacomello and M. Porfiri.
\newblock Underwater energy harvesting from a heavy flag hosting ionic polymer
  metal composites.
\newblock {\em Journal of Applied Physics}, 109(8):084903, 2011.

\bibitem{doare2011}
O. Doar{\'e} and S. Michelin.
\newblock Piezoelectric coupling in energy-harvesting fluttering flexible
  plates: linear stability analysis and conversion efficiency.
\newblock {\em Journal of Fluids and Structures}, 27(8):1357--1375, 2011.

\bibitem{michelin2013}
S. Michelin and O. Doar{\'e}.
\newblock Energy harvesting efficiency of piezoelectric flags in axial flows.
\newblock {\em Journal of Fluid Mechanics}, 714:489--504, 2013.

\bibitem{watanabe2002exp}
Y. Watanabe, S. Suzuki, M. Sugihara, and Y. Sueoka.
\newblock An experimental study of paper flutter.
\newblock {\em Journal of fluids and Structures}, 16(4):529--542, 2002.

\bibitem{watanabe2002theo}
Y. Watanabe, K. Isogai, S. Suzuki, and M. Sugihara.
\newblock A theoretical study of paper flutter.
\newblock {\em Journal of Fluids and Structures}, 16(4):543--560, 2002.

\bibitem{hidalgo2010}
P. Hidalgo, F. Herrault, A. Glezer, M. Allen, S. Kaslusky, and
  B.~St Rock.
\newblock Heat transfer enhancement in high-power heat sinks using active reed
  technology.
\newblock In {\em 2010 16th International Workshop on Thermal Investigations of
  ICs and Systems (THERMINIC)}, pages 1--6. IEEE, 2010.

\bibitem{shoele2014}
K. Shoele and R. Mittal.
\newblock Computational study of flow-induced vibration of a reed in a channel
  and effect on convective heat transfer.
\newblock {\em Physics of Fluids}, 26(12):127103, 2014.

\bibitem{huang1995}
L. Huang, S.J. Quinn, P.D.M. Ellis, and J.E. Ffowcs Williams.
\newblock Biomechanics of snoring.
\newblock {\em Endeavour}, 19(3):96--100, 1995.

\bibitem{auregan1995}
Y. Auregan and C. Depollier.
\newblock Snoring: linear stability analysis and in-vitro experiments.
\newblock {\em Journal of Sound and Vibration}, 188(1):39--53, 1995.

\bibitem{balint2005}
T.S. Balint and A.D. Lucey.
\newblock Instability of a cantilevered flexible plate in viscous channel flow.
\newblock {\em Journal of Fluids and Structures}, 20(7):893--912, 2005.

\bibitem{howell2009}
R.H. Howell, A.D. Lucey, P.W. Carpenter, and M.W. Pitman.
\newblock Interaction between a cantilevered-free flexible plate and ideal
  flow.
\newblock {\em Journal of Fluids and Structures}, 25(3):544--566, 2009.

\bibitem{shoele2016b}
K. Shoele and R. Mittal.
\newblock Flutter instability of a thin flexible plate in a channel.
\newblock {\em Journal of Fluid Mechanics}, 786:29--46, 2016.

\bibitem{wu2005}
X. Wu and S. Kaneko.
\newblock Linear and nonlinear analyses of sheet flutter induced by leakage
  flow.
\newblock {\em Journal of Fluids and Structures}, 20(7):927--948, 2005.

\bibitem{dessi2015}
D. Dessi and S. Mazzocconi.
\newblock Aeroelastic behavior of a flag in ground effect.
\newblock {\em Journal of Fluids and Structures}, 55:303--323, 2015.

\bibitem{alben2015}
S. Alben.
\newblock Flag flutter in inviscid channel flow.
\newblock {\em Physics of Fluids}, 27(3):033603, 2015.

\bibitem{doare2012}
O. Doar{\'e} and C. Eloy.
\newblock The influence of channel walls on flag flutter.
\newblock In {\em Flow Induced Vibration}, 2012.

\bibitem{guo2000}
C.Q. Guo and M.P. Paidoussis.
\newblock Stability of rectangular plates with free side-edges in
  two-dimensional inviscid channel flow.
\newblock {\em Journal of Applied Mechanics}, 67(1):171--176, 2000.

\bibitem{jaiman2014}
R.K. Jaiman, M.K. Parmar, and P.S. Gurugubelli.
\newblock Added mass and aeroelastic stability of a flexible plate interacting
  with mean flow in a confined channel.
\newblock {\em Journal of Applied Mechanics}, 81(4):041006, 2014.

\bibitem{brennen1982}
C.E. Brennen.
\newblock A review of added mass and fluid inertial forces.
\newblock Technical report, BRENNEN (CE) SIERRA MADRE CA, 1982.

\bibitem{doare2011num}
O. Doar{\'e}, M. Sauzade, and C. Eloy.
\newblock Flutter of an elastic plate in a channel flow: Confinement and
  finite-size effects.
\newblock {\em Journal of Fluids and Structures}, 27(1):76--88, 2011.

\bibitem{cisonni2017}
J. Cisonni, A.D. Lucey, N.S.J. Elliott, and M. Heil.
\newblock The stability of a flexible cantilever in viscous channel flow.
\newblock {\em Journal of Sound and Vibration}, 396:186--202, 2017.

\bibitem{doare2011exp}
O. Doar{\'e}, D. Mano, and J. C. Bilbao~Ludena.
\newblock Effect of spanwise confinement on flag flutter: Experimental
  measurements.
\newblock {\em Physics of Fluids}, 23(11):111704, 2011.

\bibitem{trasch2018}
M. Tr{\"a}sch, A. D{\'e}porte, S. Delacroix, J.-B. Drevet,
  B. Gaurier, and G. Germain.
\newblock Power estimates of an undulating membrane tidal energy converter.
\newblock {\em Ocean Engineering}, 148:115--124, 2018.

\bibitem{muriel2018}
D.F. Muriel and E.A. Cowen.
\newblock On the realization of a second buckling mode in a
  periodically-constrained heavy elastica.
\newblock {\em Extreme Mechanics Letters}, 21:76--81, 2018.

\bibitem{muriel2016}
D.F. Muriel, R.O. Tinoco, B.P. Filardo, and E.A. Cowen.
\newblock Development of a novel, robust, sustainable and low cost self-powered
  water pump for use in free-flowing liquid streams.
\newblock {\em Renewable Energy}, 91:466--476, 2016.

\bibitem{alam2012}
Mohammad-Reza Alam.
\newblock Nonlinear analysis of an actuated seafloor-mounted carpet for a
  high-performance wave energy extraction.
\newblock In {\em Proc. R. Soc. A}, volume 468, pages 3153--3171. The Royal
  Society, 2012.

\bibitem{cho2000}
IH~Cho and MH~Kim.
\newblock Interactions of horizontal porous flexible membrane with waves.
\newblock {\em Journal of waterway, port, coastal, and ocean engineering},
  126(5):245--253, 2000.

\bibitem{schouveiler2009}
Lionel Schouveiler and Christophe Eloy.
\newblock Coupled flutter of parallel plates.
\newblock {\em Physics of fluids}, 21(8):081703, 2009.

\bibitem{mougel2016}
J{\'e}r{\^o}me Mougel, Olivier Doar{\'e}, and S{\'e}bastien Michelin.
\newblock Synchronized flutter of two slender flags.
\newblock {\em Journal of Fluid Mechanics}, 801:652--669, 2016.

\bibitem{xia2015}
Yifan Xia, S{\'e}bastien Michelin, and Olivier Doar{\'e}.
\newblock Fluid-solid-electric lock-in of energy-harvesting piezoelectric
  flags.
\newblock {\em Physical Review Applied}, 3(1):014009, 2015.

\bibitem{xia2017}
Y.~Xia, O.~Doar\'e, and S.~Michelin.
\newblock Fluid-solid-electric energy transport along piezoelectric flags.
\newblock {\em Eur. J. Comp. Mech.}, 26:154--171, 2017.

\bibitem{virot2016}
E.~Virot, X.~Amandolese, and P.~Hemon.
\newblock Coupling between a flag and a spring-mass oscillator.
\newblock {\em J. Fluids Struct.}, 65:447--454, 2016.

\bibitem{shelley2005}
Michael Shelley, Nicolas Vandenberghe, and Jun Zhang.
\newblock Heavy flags undergo spontaneous oscillations in flowing water.
\newblock {\em Physical review letters}, 94(9):094302, 2005.

\bibitem{jia2007}
Lai-Bing Jia, Fang Li, Xie-Zhen Yin, and Xie-Yuan Yin.
\newblock Coupling modes between two flapping filaments.
\newblock {\em Journal of Fluid Mechanics}, 581:199--220, 2007.

\bibitem{kornecki1976}
A~Kornecki, EH~Dowell, and J~O'brien.
\newblock On the aeroelastic instability of two-dimensional panels in uniform
  incompressible flow.
\newblock {\em Journal of Sound and Vibration}, 47(2):163--178, 1976.

\bibitem{tang2008}
Liaosha Tang, Michael~P Pa{\i}, et~al.
\newblock The influence of the wake on the stability of cantilevered flexible
  plates in axial flow.
\newblock {\em Journal of Sound and Vibration}, 310(3):512--526, 2008.

\bibitem{cairns1979}
RA~Cairns.
\newblock The role of negative energy waves in some instabilities of parallel
  flows.
\newblock {\em Journal of Fluid Mechanics}, 92(1):1--14, 1979.

\bibitem{benjamin1963}
T~Brooke Benjamin.
\newblock The threefold classification of unstable disturbances in flexible
  surfaces bounding inviscid flows.
\newblock {\em Journal of Fluid Mechanics}, 16(3):436--450, 1963.

\bibitem{tophoj2013}
L~Toph{\o}j, Jerome Mougel, Tomas Bohr, and David Fabre.
\newblock Rotating polygon instability of a swirling free surface flow.
\newblock {\em Physical review letters}, 110(19):194502, 2013.

\bibitem{sakai1989}
Satoshi Sakai.
\newblock Rossby-kelvin instability: a new type of ageostrophic instability
  caused by a resonance between rossby waves and gravity waves.
\newblock {\em Journal of Fluid Mechanics}, 202:149--176, 1989.

\bibitem{joarder1997}
PS~Joarder, VM~Nakariakov, and B~Roberts.
\newblock A manifestation of negative energy waves in the solar atmosphere.
\newblock {\em Solar Physics}, 176(2):285--297, 1997.

\bibitem{lehmann2013}
Marcus Lehmann, Ryan Elandt, Henry Pham, Reza Ghorbani, Mostafa Shakeri, and
  Mohammad-Reza Alam.
\newblock An artificial seabed carpet for multidirectional and broadband wave
  energy extraction: Theory and experiment.
\newblock In {\em Proceedings of 10th European Wave and Tidal Energy
  Conference, EWTEC2013, Aalborg, Denmark}, 2013.

\end{thebibliography}

\end{document}